\documentclass{article}
\usepackage{graphicx} % Required for inserting images
\usepackage[T1]{fontenc}
\usepackage[english]{babel}
\usepackage[utf8]{inputenc}
\usepackage{fancyhdr}
\usepackage{amsmath}
\usepackage{physics}
\usepackage{amssymb}
\usepackage{subcaption}
\usepackage{float}
\usepackage{caption}
\graphicspath{ "C:\Users\kdvpo\Desktop\QuaternionisWannier\images" }

\usepackage[margin=1in,footskip=0.3in]{geometry}
\usepackage{wrapfig}
\usepackage{hyperref}
\usepackage{braket}
\usepackage{mathtools}
\usepackage{cite}

\usepackage{authblk}
\title{Quaternionic description of semiconductor position-based qubits}
\author[1,2]{Wojciech Nowakowski}
\author[2]{Krzysztof Pomorski}
\affil[1]{Lodz University of Technology, Institute of Physics, Lodz, Poland }
\affil[2]{Quantum Hardware Systems, Lodz, Poland  }
\date{}

\begin{document}
\maketitle
\begin{abstract}
    The quaternionic description of semiconductor single-electron devices is given in the single-electron regime.
    %This paper contains the results of work aimed at solving the problem of the quaternionic description of one-electron devices such as the quantum dot.
    %The aim was to quaternionise the Hamiltonians describing such devices. Both the case of a single dot and of two interacting dots were considered. 
    The conversion scheme of complex value Hamiltonian into a quaternion is formulated for the case of single-electron semiconductor qubit and many electrostatically interacting qubits.
    %, rules for the inverse conversion, and explains the course of considerations.
    In particular, the quantum evolution operator is presented in quaternion form for the case of one and many electrostatically interacting quantum bodies. 
\end{abstract}

%\begin{keyword}Something 
%    Something else
%\end{keyword}

\section{Introduction}
\begin{figure}
    \centering
    \begin{subfigure}{\textwidth}
         \centering
         \includegraphics[width=\textwidth]{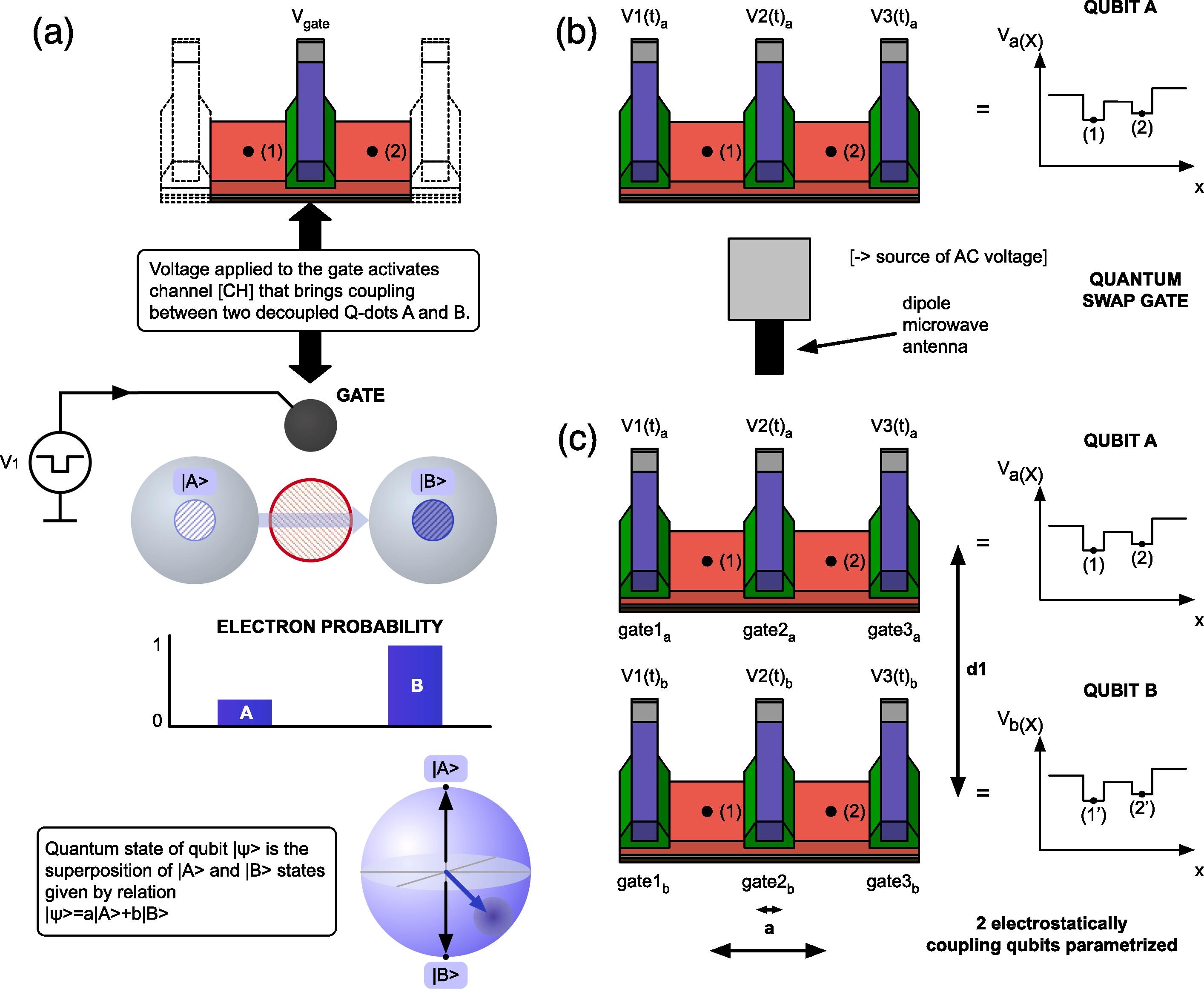}
         \caption{Concept of position based qubit and quantum logic gate made from position based qubits as by \cite{Pomorski_2019} .}
         \label{fig:FIG1}
     \end{subfigure}
     \hfill
     \begin{subfigure}{\textwidth}
         \centering
         \includegraphics[width=0.6\textwidth]{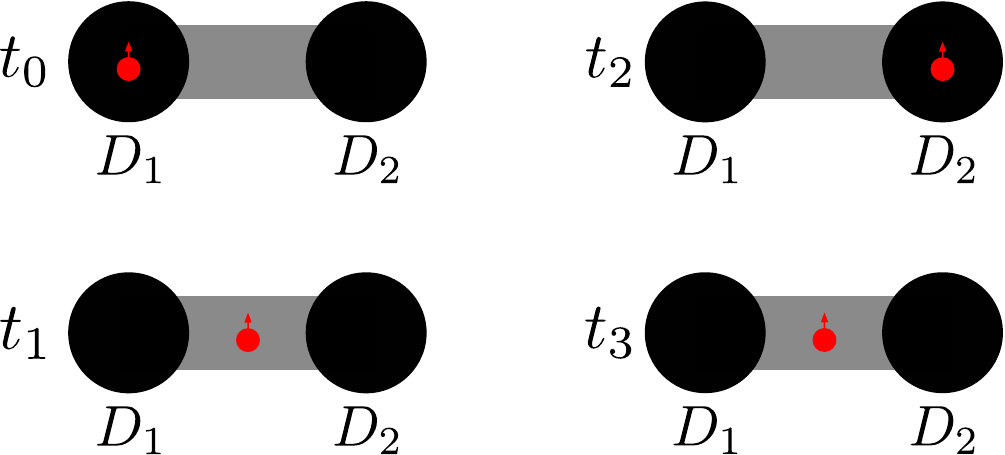}
         \caption{Electron occupancy in position based qubit at different time instances, when $t_1<t_2<t_3<t_4$.}
         \label{fig:IntQD}
     \end{subfigure}
        \caption{Conceptual description of position-based qubits in semiconductor quantum dots as given by \cite{Cryogenics}.}
        \label{fig:Pic}
\end{figure}

The quaternionic representation of various sets of differential equations can be very compact and useful \cite{QutMax}, as is the case of Maxwell differential equations, where four equations can be represented by one.
Quaternions are used in quantum mechanics to describe Schrödinger's equation \cite{Quantum_field},\cite{e22121424}, Stone's quantum theorem \cite{10.1063/1.1703794}, and annihilation-creation operators \cite{HORWITZ1984432}. However, no such description exists for adjacent semiconductor quantum dots that implement position-based qubits such as those depicted in Fig.\ref{fig:IntQD}.
%Based on the work on interacting single-electrode devices \cite{Pomorski_2019}, we have developed a quaternionic description of this type of device.
In the following sections, we present the quaternion description of single-electron devices, based on the work \cite{Pomorski_2019}.

\section{Quaternionic view of two interacting position-based qubits}
\subsection{Tight-binding model of position-based qubit}
Hamiltonian of tight-binding model describing single-electron devices(as presented by Fujisawa \cite{FUJISAWA20041046} or Pomorski \cite{ma17194846}) uses Hamiltonian of position-based qubit $\hat{H}_P=\hat{R} \hat{H}_E \hat{R}^{-1}$ with rotation matrix $\hat{R}$, fictitious angle $\Theta_s$ and eigenenergy Hamiltonian $\hat{H}_E$. That in case of two energy level system can be expressed in the following form:
\begin{multline}
        \label{eq:Hp}
    \hat{H}_P = \begin{bmatrix}
        E_{P1} & t_s\cos{\theta_P}+\text{i}t_s \sin{\theta_P} \\
        t_s\cos{\theta_P}-\text{i}t_s\sin{\theta_P} & E_{P2}
    \end{bmatrix}
    =
    \begin{bmatrix}
        E_{P1} & t_s e^{+i\theta_P} \\
        t_se^{-i\theta_P} & E_{P2}
    \end{bmatrix} = \hat{R}[ E_1 \ket{E_1}\bra{E_1}+E_2 \ket{E_2}\bra{E_2} ] \hat{R}^{-1} \\ 
    =
    \begin{bmatrix}
        +\cos(\Theta_s) & +\sin(\Theta_s) \\
        -\sin(\Theta_s) & +\cos(\Theta_s)
    \end{bmatrix}
    \begin{bmatrix}
        E_{1} & 0 \\
        0 & E_{2}
    \end{bmatrix}
    \begin{bmatrix}
        +\cos(\Theta_s) & -\sin(\Theta_s) \\
        +\sin(\Theta_s) & +\cos(\Theta_s)
    \end{bmatrix}= \hat{R}\hat{H}_E \hat{R}^{-1}= \\
    =
     \begin{bmatrix}
        E_{1}\cos(\Theta_s)^2+E_{2}\sin(\Theta_s)^2 & (E_{2}-E_{1})\sin(\Theta_s)\cos(\Theta_s) \\
        +(E_{2}-E_{1})\sin(\Theta_s)\cos(\Theta_s)  & E_{2}\cos(\Theta_s)^2+E_{1}\sin(\Theta_s)^2
    \end{bmatrix}
    =
    \begin{bmatrix}
        E_{1}+(E_{2}-E_{1})\sin(\Theta_s)^2 & \frac{1}{2}(E_{2}-E_{1})\sin(2\Theta_s) \\
        +\frac{1}{2}(E_{2}-E_{1})\sin(2\Theta_s) & (E_{2}-E_{1})\cos(\Theta_s)^2+E_{1}
    \end{bmatrix}=   \\
    =
      \begin{bmatrix}
        \frac{1}{2}(E_{2}+E_{1}) & 0 \\
        0 & \frac{1}{2}(E_{2}+E_{1})
    \end{bmatrix} +
      \begin{bmatrix}
         E_{1}\cos(\Theta_s)^2+E_{2}\sin(\Theta_s)^2-\frac{E_1+E_2}{2} & \frac{1}{2}(E_{2}-E_{1})\sin(2\Theta_s) \\
        +\frac{1}{2}(E_{2}-E_{1})\sin(2\Theta_s) &   E_{1}\sin(\Theta_s)^2+E_{2}\cos(\Theta_s)^2-\frac{E_1+E_2}{2}
    \end{bmatrix}=
     \\
     \frac{E_1+E_2}{2}\hat{I}
     +
     \begin{bmatrix}
         E_{1}(1-\sin(\Theta_s)^2)+E_{2}\sin(\Theta_s)^2-\frac{E_1+E_2}{2} & \frac{1}{2}(E_{2}-E_{1})\sin(2\Theta_s) \\
        +\frac{1}{2}(E_{2}-E_{1})\sin(2\Theta_s) &   E_{1}\sin(\Theta_s)^2+E_{2}(1-\sin(\Theta_s))^2-\frac{E_1+E_2}{2}
    \end{bmatrix}        .
\end{multline}
We continue elementary mathematical steps 
\begin{multline}
\hat{H}_P=
 \begin{bmatrix}
         E_{1}(1-\sin(\Theta_s)^2)+E_{2}\sin(\Theta_s)^2-\frac{E_1+E_2}{2} & \frac{1}{2}(E_{2}-E_{1})\sin(2\Theta_s) \\
        +\frac{1}{2}(E_{2}-E_{1})\sin(2\Theta_s) &   E_{1}\sin(\Theta_s)^2+E_{2}(1-\sin(\Theta_s))^2-\frac{E_1+E_2}{2}
    \end{bmatrix} + \frac{1}{2}(E_{2}+E_{1})\hat{I} = \nonumber \\ =
    \frac{1}{2}(E_{2}+E_{1})\hat{I}\hat{\sigma}_0 
    + \frac{1}{2}(E_{2}-E_{1})\sin(2\Theta_s)
    \hat{\sigma}_1 + \Big[ (E_{2}-E_{1})\sin(\Theta_s)^2-\frac{E_2-E_1}{2} \Big] \hat{\sigma}_3= 
     \\
     =
    \begin{bmatrix}
         +[(E_{2}-E_{1})\sin(\Theta_s)^2-\frac{E_2-E_1}{2}] & \frac{1}{2}(E_{2}-E_{1})\sin(2\Theta_s) \\
        +\frac{1}{2}(E_{2}-E_{1})\sin(2\Theta_s) &   -[(E_{2}-E_{1})\sin(\Theta_s)^2-\frac{E_2-E_1}{2}]
    \end{bmatrix}+ \frac{1}{2}(E_{2}+E_{1})\hat{I} = \nonumber \\ 
    %\\=  \\
    =(E_{2}-E_{1})
     \begin{bmatrix}
         +[\sin(\Theta_s)^2-\frac{1}{2}] & \frac{1}{2}\sin(2\Theta_s) \\
        +\frac{1}{2}\sin(2\Theta_s) &   -[\sin(\Theta_s)^2-\frac{1}{2}]
    \end{bmatrix}+ \frac{1}{2}(E_{2}+E_{1})\hat{I}= \nonumber \\
   = (E_{2}-E_{1})
     \begin{bmatrix}
         +[\sin(\Theta_s)^2-\frac{1}{2}] & \pm \sin(\Theta_s)\sqrt{1-\sin(\Theta_s)^2} \\
        \pm \sin(\Theta_s)\sqrt{1-\sin(\Theta_s)^2} &   -[\sin(\Theta_s)^2-\frac{1}{2}]
    \end{bmatrix}+ \frac{1}{2}(E_{2}+E_{1})\hat{I} = \nonumber \\
    = (E_{2}-E_{1})\sin(\Theta_s)^2
     \begin{bmatrix}
         +[1-\frac{1}{2\sin(\Theta_s)^2}] & \pm \sqrt{\frac{1}{\sin(\Theta_s)^2}-1} \\
        \pm \sqrt{\frac{1}{\sin(\Theta_s)^2}-1} &   -[1-\frac{1}{2\sin(\Theta_s)^2}]
    \end{bmatrix}+ \frac{1}{2}(E_{2}+E_{1})\hat{I} 
    = \nonumber \\
    = \frac{(E_{2}-E_{1})}{2}\frac{1}{w}
     \begin{bmatrix}
         +[1-w] & \pm \sqrt{2}\sqrt{w-\frac{1}{2}} \\
        \pm \sqrt{2}\sqrt{w-\frac{1}{2}} &   -[1-w]
    \end{bmatrix}+ \frac{1}{2}(E_{2}+E_{1})\hat{I},
    w=\frac{1}{2 \sin(\Theta_s)^2}, 
    \nonumber \\
\end{multline}
where we have used the property of maximization of
the presence of the left Wannier wavefunction $w_L(x)=\cos(\Theta_s)\psi_{E1}(x)+\sin(\Theta_s)\psi_{E2}(x)$ in the region from $-\infty$ to $x_0$ encoded in minimization of functional $\int_{-\infty}^{x_0}|w_L(x)|^2dx=F[\Theta_s]$, so it is achieved that $\frac{d}{d\Theta_s}F[\Theta_s]=0$. 
Having the logical state zero assigned from $-\infty$ to $x_0$ (medium distance between 2 quantum dots)
we obtain the formula for angle $\Theta_s$ as given by \cite{ma17194846} and the expression
\begin{eqnarray}
      \Theta_s=\frac{1}{2}\arctan{\Bigg[ \frac{\int_{-\infty}^{+x_0}\frac{1}{2}[\psi_{E1}(x)\psi_{E2}^{*}(x)+\psi_{E1}^{*}(x)\psi_{E2}(x)]dx}{\int_{-\infty}^{+x_0}dx[|\psi_{E2}(x)|^2-|\psi_{E1}(x)|^2]} \Bigg]}.
  \end{eqnarray}
  It is useful to parameterize $\hat{H}_P$ by the introduced quantity $w=\frac{1}{2 \sin(\Theta_s)^2}$ and two eigenenergies $E_1$ and $E_2$. It is proper to underline that 
 \begin{eqnarray}
    \hat{H}_E=
     \begin{bmatrix}
        [ - \frac{\hbar^2}{2m}\frac{d^2}{dx^2} + V_{eff}(x)  ] & 0 \\
        0 & [ - \frac{\hbar^2}{2m}\frac{d^2}{dx^2} + V_{eff}(x)  ]
    \end{bmatrix}
    =
     \begin{bmatrix}
        E_{1} & 0 \\
        0 & E_{2}
    \end{bmatrix}=E_1\ket{E_1}\bra{E_1}+E_2\ket{E_2}\bra{E_2}, \nonumber \\
      \hat{H}_E \ket{\psi_{E}}=
     \hat{H}_E 
     \begin{bmatrix}
       \sqrt{p_{E1}(t)}e^{i \gamma_1(t)}\psi_{E1}(x) \\
         \sqrt{p_{E2}(t)}e^{i \gamma_2(t)}
         \psi_{E2}(x)
    \end{bmatrix} = i \hbar \frac{d}{dt}
    \begin{bmatrix}
       \sqrt{p_{E1}(t)}e^{i \gamma_1(t)}\psi_{E1}(x) \\
         \sqrt{p_{E2}(t)}e^{i \gamma_2(t)}\psi_{E2}(x)
    \end{bmatrix} = i \hbar \frac{d}{dt}\ket{\psi_{E}}= \nonumber \\
    =
    \begin{bmatrix}
        +\cos(\Theta_s) & -\sin(\Theta_s) \\
        +\sin(\Theta_s) & +\cos(\Theta_s)
    \end{bmatrix}
     \begin{bmatrix}
        +\cos(\Theta_s) & +\sin(\Theta_s) \\
        -\sin(\Theta_s) & +\cos(\Theta_s)
    \end{bmatrix}i \hbar \frac{d}{dt}
    \begin{bmatrix}
       \sqrt{p_{E1}(t)}e^{i \gamma_1(t)}\psi_{E1}(x) \\
         \sqrt{p_{E2}(t)}e^{i \gamma_2(t)}\psi_{E2}(x)
    \end{bmatrix}, \nonumber \\
    \ket{\psi}_P=
    \begin{bmatrix}
       w_L(x) \\
       w_R(x)
    \end{bmatrix}=
    \begin{bmatrix}
        +\cos(\Theta_s) & +\sin(\Theta_s) \\
        -\sin(\Theta_s) & +\cos(\Theta_s)
    \end{bmatrix}
    \begin{bmatrix}
      \psi_{E1}(x) \\
      \psi_{E2}(x)
    \end{bmatrix}, %\nonumber \\
     \ket{\psi}_P=
    \begin{bmatrix}
       \psi_{E1}(x) \\
       \psi_{E2}(x)
    \end{bmatrix}=
    \begin{bmatrix}
        +\cos(\Theta_s) & -\sin(\Theta_s) \\
        +\sin(\Theta_s) & +\cos(\Theta_s)
    \end{bmatrix}
    \begin{bmatrix}
      w_L(x) \\
      w_R(x)
    \end{bmatrix}, \nonumber \\
     \hat{H}_E \ket{\psi}_E=
    \hat{H}_E 
     \begin{bmatrix}
       \sqrt{p_{E1}(t)}e^{i \gamma_1(t)}\psi_{E1}(x) \\
         \sqrt{p_{E2}(t)}e^{i \gamma_2(t)}
         \psi_{E2}(x)
    \end{bmatrix} =  \hat{H}_E 
    \begin{bmatrix}
        \sqrt{p_{E1}(t)}e^{i \gamma_1(t)} & 0 \\
        0 & \sqrt{p_{E2}(t)}e^{i \gamma_2(t)}
    \end{bmatrix}
      \begin{bmatrix}
        +\cos(\Theta_s) & -\sin(\Theta_s) \\
        +\sin(\Theta_s) & +\cos(\Theta_s)
    \end{bmatrix}
    \begin{bmatrix}
      w_L(x) \\
      w_R(x)
    \end{bmatrix}=
    \nonumber \\
    i \hbar \frac{d}{dt} \Bigg[
    \sin(\Theta_s)[ \sqrt{p_{E2}(t)}e^{i \gamma_2(t)}-\sqrt{p_{E1}(t)}e^{i \gamma_1(t)} ]
     \begin{bmatrix}
       0 &  1\\
       1 & 0
    \end{bmatrix}
    + \nonumber \\
    +
    \begin{bmatrix}
        \cos(\Theta_s) & -\sin(\Theta_s) \\
        \sin(\Theta_s) & +\cos(\Theta_s)
    \end{bmatrix}
    \begin{bmatrix}
        \sqrt{p_{E1}(t)}e^{i \gamma_1(t)} & 0 \\
        0 & \sqrt{p_{E2}(t)}e^{i \gamma_2(t)}
    \end{bmatrix} \Bigg]
    \begin{bmatrix}
      w_L(x) \\
      w_R(x)
    \end{bmatrix}.
\end{eqnarray}
Consequently using the fact that $\Theta_s$ is time independent we obtain the following chain of equivalent expressions as giving the transformation from energy base to the position base
\begin{eqnarray}
 \hat{H}_E 
  \begin{bmatrix}
        +\cos(\Theta_s) & -\sin(\Theta_s) \\
        +\sin(\Theta_s) & +\cos(\Theta_s)
    \end{bmatrix}
    \begin{bmatrix}
        +\cos(\Theta_s) & +\sin(\Theta_s) \\
        +\sin(\Theta_s) & -\cos(\Theta_s)
    \end{bmatrix} 
    \times \nonumber \\
    \times 
    \begin{bmatrix}
        \sqrt{p_{E1}(t)}e^{i \gamma_1(t)} & 0 \\
        0 & \sqrt{p_{E2}(t)}e^{i \gamma_2(t)}
    \end{bmatrix}
      \begin{bmatrix}
        +\cos(\Theta_s) & -\sin(\Theta_s) \\
        +\sin(\Theta_s) & +\cos(\Theta_s)
    \end{bmatrix}
    \begin{bmatrix}
      w_L(x) \\
      w_R(x)
    \end{bmatrix}=
    \nonumber \\
    =
    \begin{bmatrix}
        +\cos(\Theta_s) & -\sin(\Theta_s) \\
        +\sin(\Theta_s) & +\cos(\Theta_s)
    \end{bmatrix} 
    \times \nonumber \\
    \times
    i\hbar \frac{d}{dt}    \Bigg[
     \begin{bmatrix}
        +\cos(\Theta_s) & +\sin(\Theta_s) \\
        +\sin(\Theta_s) & -\cos(\Theta_s)
    \end{bmatrix}
    \begin{bmatrix}
    \sqrt{p_{E1}(t)}e^{i \gamma_1(t)} &   0                                \\
    0                                 &  \sqrt{p_{E2}(t)}e^{i \gamma_2(t)}
    \end{bmatrix} 
   \begin{bmatrix}
        \cos(\Theta_s) & -\sin(\Theta_s) \\
        \sin(\Theta_s) & +\cos(\Theta_s)
    \end{bmatrix}
    \Bigg]
    \begin{bmatrix}
      w_L(x) \\
      w_R(x)
    \end{bmatrix}, \nonumber \\
    \Bigg[
    \begin{bmatrix}
        +\cos(\Theta_s) & +\sin(\Theta_s) \\
        -\sin(\Theta_s) & +\cos(\Theta_s)
    \end{bmatrix}
    \hat{H}_E 
  \begin{bmatrix}
        +\cos(\Theta_s) & -\sin(\Theta_s) \\
        +\sin(\Theta_s) & +\cos(\Theta_s)
    \end{bmatrix}  \Bigg] \times \nonumber \\ \times
    \Bigg[
     \begin{bmatrix}
        +\cos(\Theta_s) & +\sin(\Theta_s) \\
        +\sin(\Theta_s) & -\cos(\Theta_s)
    \end{bmatrix}
    \begin{bmatrix}
        \sqrt{p_{E1}(t)}e^{i \gamma_1(t)} & 0 \\
        0 & \sqrt{p_{E2}(t)}e^{i \gamma_2(t)}
    \end{bmatrix}
      \begin{bmatrix}
        +\cos(\Theta_s) & -\sin(\Theta_s) \\
        +\sin(\Theta_s) & +\cos(\Theta_s)
    \end{bmatrix} \Bigg]
    \begin{bmatrix}
      w_L(x) \\
      w_R(x)
    \end{bmatrix}=
    \nonumber \\
  i\hbar \frac{d}{dt}    \Bigg[
     \begin{bmatrix}
        +\cos(\Theta_s) & +\sin(\Theta_s) \\
        +\sin(\Theta_s) & -\cos(\Theta_s)
    \end{bmatrix}
    \begin{bmatrix}
    \sqrt{p_{E1}(t)}e^{i \gamma_1(t)} & 0 \\
        0 &  \sqrt{p_{E2}(t)}e^{i \gamma_2(t)}
    \end{bmatrix} 
    \begin{bmatrix}
        \cos(\Theta_s) & -\sin(\Theta_s) \\
        \sin(\Theta_s) & +\cos(\Theta_s)
    \end{bmatrix}
    \Bigg]
    \begin{bmatrix}
      w_L(x) \\
      w_R(x)
    \end{bmatrix}= \nonumber \\
    i\hbar \frac{d}{dt}    
    \begin{bmatrix}
    \sqrt{p_{E1}(t)}e^{i \gamma_1(t)}\cos(\Theta_s)^2+\sqrt{p_{E2}(t)}e^{i \gamma_2(t)}\sin(\Theta_s)^2 & [ \sqrt{p_{E2}(t)}e^{i \gamma_2(t)} - \sqrt{p_{E1}(t)}e^{i \gamma_1(t)}]\sin(\Theta_s)\cos(\Theta_s) \\
         [ \sqrt{p_{E2}(t)}e^{i \gamma_2(t)} - \sqrt{p_{E1}(t)}e^{i \gamma_1(t)}]\sin(\Theta_s)\cos(\Theta_s) &  \sqrt{p_{E1}(t)}e^{i \gamma_1(t)}\sin(\Theta_s)^2+\sqrt{p_{E2}(t)}e^{i \gamma_2(t)}\cos(\Theta_s)^2
    \end{bmatrix} 
    \begin{bmatrix}
      w_L(x) \\
      w_R(x)
    \end{bmatrix}= \nonumber \\
    = i \hbar  \frac{d}{dt}\ket{\psi}_P=\hat{H}_P\ket{\psi}_P= \nonumber \\
      i\hbar \frac{d}{dt}    
     \begin{bmatrix}
        +\cos(\Theta_s) & +\sin(\Theta_s) \\
        -\sin(\Theta_s) & +\cos(\Theta_s)
    \end{bmatrix}
    \begin{bmatrix}
    \sqrt{p_{E1}(t)}e^{i \gamma_1(t)} & 0 \\
        0 &  \sqrt{p_{E2}(t)}e^{i \gamma_2(t)}
    \end{bmatrix} 
    \begin{bmatrix}
      \psi_{E1}(x) \\
      \psi_{E2}(x)
    \end{bmatrix}, \nonumber \\
    i \hbar \frac{d}{dt} \psi_p(x,t)=
   \begin{bmatrix}
      1, & 1 \\
    \end{bmatrix}
    i \hbar \frac{d}{dt}
      \ket{\psi}_P=   \begin{bmatrix}
      1, & 1 \\
    \end{bmatrix} 
    \hat{H}_P
      \ket{\psi}_P= \nonumber \\
      =
    \begin{bmatrix}
      1, & 1 \\
    \end{bmatrix} 
     \Bigg[
    \begin{bmatrix}
        +\cos(\Theta_s) & +\sin(\Theta_s) \\
        -\sin(\Theta_s) & +\cos(\Theta_s)
    \end{bmatrix}
    \hat{H}_E 
  \begin{bmatrix}
        +\cos(\Theta_s) & -\sin(\Theta_s) \\
        +\sin(\Theta_s) & +\cos(\Theta_s)
    \end{bmatrix}  \Bigg] \times \nonumber \\ \times
    \Bigg[
     \begin{bmatrix}
        +\cos(\Theta_s) & +\sin(\Theta_s) \\
        +\sin(\Theta_s) & -\cos(\Theta_s)
    \end{bmatrix}
    \begin{bmatrix}
        \sqrt{p_{E1}(t)}e^{i \gamma_1(t)} & 0 \\
        0 & \sqrt{p_{E2}(t)}e^{i \gamma_2(t)}
    \end{bmatrix}
      \begin{bmatrix}
        +\cos(\Theta_s) & -\sin(\Theta_s) \\
        +\sin(\Theta_s) & +\cos(\Theta_s)
    \end{bmatrix} \Bigg]
    \begin{bmatrix}
      w_L(x) \\
      w_R(x)
    \end{bmatrix}= \nonumber \\
     =
    \begin{bmatrix}
      1, & 1 \\
    \end{bmatrix} 
     \Bigg[
    \begin{bmatrix}
        +\cos(\Theta_s) & +\sin(\Theta_s) \\
        -\sin(\Theta_s) & +\cos(\Theta_s)
    \end{bmatrix} \nonumber \times \\
    \times
     \begin{bmatrix}
        \int_{-\infty}^{+\infty}dx_1\psi_{E1}^{*}(x_1)[-\frac{\hbar^2}{2m}\frac{d^2}{dx_1^2}+V(x_1,t)]\psi_{E1}(x_1) & \int_{-\infty}^{+\infty}dx_1\psi_{E1}(x_1)[-\frac{\hbar^2}{2m}\frac{d^2}{dx_1^2}+V(x_1,t)]\psi_{E2}^{*}(x_1) \\
        \int_{-\infty}^{+\infty}dx_1\psi_{E2}(x_1)[-\frac{\hbar^2}{2m}\frac{d^2}{dx_1^2}+V(x_1,t)]\psi_{E1}^{*}(x_1)  &  \int_{-\infty}^{+\infty}dx_1\psi_{E2}^{*}(x_1)[-\frac{\hbar^2}{2m}\frac{d^2}{dx^2}+V(x,t)]\psi_{E2}(x_1)
    \end{bmatrix} \times \nonumber \\
    \times
  \begin{bmatrix}
        +\cos(\Theta_s) & -\sin(\Theta_s) \\
        +\sin(\Theta_s) & +\cos(\Theta_s)
    \end{bmatrix}  \Bigg] \times \nonumber \\ \times
    \Bigg[
     \begin{bmatrix}
        +\cos(\Theta_s) & +\sin(\Theta_s) \\
        +\sin(\Theta_s) & -\cos(\Theta_s)
    \end{bmatrix}
    \begin{bmatrix}
        \sqrt{p_{E1}(t)}e^{i \gamma_1(t)} & 0 \\
        0 & \sqrt{p_{E2}(t)}e^{i \gamma_2(t)}
    \end{bmatrix}
\Bigg]
    \begin{bmatrix}
      \psi_{E1}(x) \\
      \psi_{E2}(x)
    \end{bmatrix}, \nonumber %\\
\end{eqnarray}
where we have introduced the complex value function $\psi_P(x)$ expressed as a linear combination of maximal localized Wannier functions localized on the left or on the right side ($w_L(x)$ and $w_R(x)$) or as linear combination of 2 lowest energy eigenstates $\psi_{E1}(x)$ and $\psi_{E2}(x)$ with effective potential $V(x,t)$ describing a position-based qubit in the single electron regime as depicted in Fig. \ref{fig:Pic}. We assume that this potential is only weakly time-dependent. 
%%%%%%%%%%%% Beginnig of the problem
\begin{eqnarray}
\hat{H}_p=
\left(
\begin{array}{cc}
 \sin (2\Theta_s)  \frac{E_{12}+E_{21}}{2}+E_{1} \cos ^2(\Theta_s)+E_{2} \sin ^2(\Theta_s) & E_{12} \cos ^2(\Theta_s)+\sin
   (2\Theta_s) \frac{E_{2}-E_{1}}{2}-E_{21} \sin ^2(\Theta_s) \\
 -E_{12} \sin ^2(\Theta_s)+\sin(2\Theta_s)  \frac{E_2-E_1}{2}+E_{21} \cos ^2(\Theta_s) & -\sin (2\Theta_s)
\frac{E_{12}+E_{21}}{2}+E_1 \sin ^2(\Theta_s)+E_2 \cos ^2(\Theta_s) \\
\end{array}
\right), \nonumber \\
E_{21}(t)=E_{12}^{*}(t)=\int_{-\infty}^{-\infty}dx_1
\psi_{E2}^{*}(x_1,t)[-\frac{\hbar^2}{2m}\frac{d^2}{dx^2}+V(x_1,t)]\psi_{E2}(x_1,t), \nonumber \\
E_{ss}(t)=E_{s}(t)=\int_{-\infty}^{-\infty}dx_1
\psi_{Es}^{*}(x_1,t)[-\frac{\hbar^2}{2m}\frac{d^2}{dx^2}+V(x_1,t)]\psi_{Es}(x_1,t).
\end{eqnarray}
%%%%%%%%%%%% End of the problem

We observe that a given quantum system expressing single-electron qubits can be described by eigenenergy Hamiltonian $H_{E}$ and this picture can be transferred into position-based Hamiltonian $H_{P}$ or reversely. Moving back and forth can be achieved with the sequence of operations given in the following, where the quantum density matrix has also been introduced in the own-reality picture
$ \hat{\rho}_E = \ket{\psi_E}\bra{\psi_E} $ or in the position-based picture
$ \hat{\rho}_P = \ket{\psi_P}\bra{\psi_P} $. 
\begin{eqnarray}
\ket{\psi}_{P}
     \rightarrow
    \ket{\psi}_{E}=\hat{R}^{-1} \ket{\psi}_{P} 
    =
    \begin{bmatrix}
        \sqrt{p_{E1}(t)}e^{i \gamma_1(t)} & 0 \\
        0 & \sqrt{p_{E2}(t)}e^{i \gamma_2(t)}
    \end{bmatrix} 
    \begin{bmatrix}
      \psi_{E1}(x) \\
      \psi_{E2}(x)
    \end{bmatrix} =
    \Bigg[
    \begin{bmatrix}
        \sqrt{p_{E1}}e^{i \gamma_1} & 0 \\
        0 & \sqrt{p_{E2}}e^{i \gamma_2}
    \end{bmatrix} 
     \hat{R} \Bigg]
    \begin{bmatrix}
      w_{L}(x) \\
      w_{R}(x)
    \end{bmatrix}, \nonumber \\
    \Bigg[
   \ket{\psi}_E \rightarrow \ket{\psi}_P, 
   \begin{bmatrix}
       \psi_{E1}(x) \\
       \psi_{E2}(x)
    \end{bmatrix}
    \rightarrow 
     \begin{bmatrix}
       w_L(x) \\
       w_R(x)
    \end{bmatrix}
    \Bigg] :
    \begin{bmatrix}
        \sqrt{p_{E1}(t)}e^{i \gamma_1(t)} & 0 \\
        0 & \sqrt{p_{E2}(t)}e^{i \gamma_2(t)}
       \end{bmatrix} \rightarrow 
       \Bigg[
      \hat{R}
     \begin{bmatrix}
         \sqrt{p_{E1}}e^{i \gamma_1} & 0 \\
         0 & \sqrt{p_{E2}}e^{i \gamma_2}
     \end{bmatrix} 
      \hat{R}^{-1} \Bigg], \nonumber \\
      ( \hat{\rho}_E=\ket{\psi_E}\bra{\psi_E} )\rightarrow 
      ( \hat{\rho}_P= \ket{\psi_P}\bra{\psi_P} 
      =
     \hat{R}\ket{\psi_E}\bra{\psi_E}\hat{R}^{-1} =
     \hat{R}\ket{\psi_E}\bra{\psi_E}\hat{R}^{T} = \hat{R} \hat{\rho}_E \hat{R}^{T} = \hat{R} \hat{\rho}_E \hat{R}^{-1}
     )  ,  \nonumber \\
      \hat{\rho}_E= \nonumber \\
      \begin{bmatrix}
        \sqrt{p_{E1}}e^{i \gamma_1} \psi_{E1}(x)  \\
        \sqrt{p_{E2}}e^{i \gamma_2} \psi_{E2}(x)
    \end{bmatrix} 
     \begin{bmatrix}
        \sqrt{p_{E1}}e^{-i \gamma_1}\psi_{E1}^{*}(x)  &  \sqrt{p_{E2}}e^{-i \gamma_2}\psi_{E2}^{*}(x)
    \end{bmatrix} =
     \begin{bmatrix}
        p_{E1} |\psi_{E1}(x)|^2 &  \sqrt{p_{E1}}\sqrt{p_{E2}}e^{i (\gamma_1-\gamma_2)}   \\
       \sqrt{p_{E1}}\sqrt{p_{E2}}e^{-i(\gamma_1-\gamma_2)}    & p_{E2} |\psi_{E2}(x)|^2
    \end{bmatrix} = \nonumber \\
     \begin{bmatrix}
        \sqrt{p_{E1}(t)}e^{i \gamma_1(t)} & 0 \\
        0 & \sqrt{p_{E2}(t)}e^{i \gamma_2(t)}
    \end{bmatrix}
      \begin{bmatrix}
        +\cos(\Theta_s) & -\sin(\Theta_s) \\
        +\sin(\Theta_s) & +\cos(\Theta_s)
    \end{bmatrix}
    \begin{bmatrix}
     | w_L(x)|^2 &  w_L(x)w_R(x)^{*} \\
       w_R(x)w_L(x)^{*}          &  |w_R(x)|^2
    \end{bmatrix} \times \nonumber \\
    \times
      \begin{bmatrix}
        +\cos(\Theta_s) & +\sin(\Theta_s) \\
        -\sin(\Theta_s) & +\cos(\Theta_s)
    \end{bmatrix}
     \begin{bmatrix}
        \sqrt{p_{E1}(t)}e^{-i \gamma_1(t)} & 0 \\
        0 & \sqrt{p_{E2}(t)}e^{-i \gamma_2(t)}
    \end{bmatrix} = \nonumber \\
    diag \Bigg[ p_{E1} \left( |w_L|^2 \cos^2(\Theta_s)+\sin(2\Theta_s)\frac{( w_L w_R^{*} + w_R w_L^{*})}{2}+|w_R|^2 \sin ^2(\Theta_s)\right), \nonumber \\
   p_{E2} \left( |w_L|^2 \sin
   ^2(\Theta_s)-\sin(2\Theta_s) \frac{( w_L w_R^{*}+ w_R w_L^{*} )}{2} + |w_R|^2 \cos ^2(\Theta_s)\right) \Bigg] + \nonumber \\
+adiag[ \sqrt{ p_{E1}}\sqrt{p_{E2}} e^{-i (\gamma_{1}-\gamma_{2})} \left(\sin (\Theta_s) \cos (\Theta_s))
   ( |w_{R}|^2-|w_L|^{2})-w_L w_R^{*} \sin ^2(\Theta_s)+ w_R w_L^{*} \cos ^2(\Theta_s)\right) ,  \nonumber \\
   \sqrt{p_{E1}} \sqrt{p_{E2}} e^{i (\gamma_{1}-\gamma_{2})} \left(\sin (\Theta_s)
   \frac{(|w_R|^{2}-|w_L|^{2})}{2}+ w_{L} w_R^{*} \cos ^2(\Theta_s)-w_R w_L^{*} \sin ^2(\Theta_s)\right)  ] = \nonumber \\
      ( \hat{\rho}_P=\ket{\psi_P}\bra{\psi_P} )\rightarrow 
     ( \hat{\rho}_E= \ket{\psi_E}\bra{\psi_E} = \nonumber \\
     =
     \hat{R}^{-1}\ket{\psi_P}\bra{\psi_P}\hat{R}=
     \hat{R}^{T}\ket{\psi_P}\bra{\psi_P}\hat{R}=
     \hat{R}^{T}\hat{\rho}_P\hat{R}
     =\hat{R}^{-1}\hat{\rho}_P\hat{R}, \nonumber \\
     i \hbar \frac{d}{dt}\rho_E = [ \hat{H}_E , \hat{\rho}_E ], 
     i \hbar \frac{d}{dt}\rho_P = [ \hat{H}_P , \hat{\rho}_P ]
     , \nonumber \\
     \int_{-\infty}^{+\infty}dx \psi_{E_k}^{*}(x) \psi_{E_s}(x) =\delta_{s,k}, 
     \int_{-\infty}^{+\infty}dx w_L^{*}(x)w_R(x)=0,  \int_{-\infty}^{+\infty}dx w_L^{*}(x)w_L(x)=\int_{-\infty}^{+\infty}dx w_R^{*}(x)w_R(x)=1, \nonumber \\
     \hat{\rho}_E(t)=e^{\frac{\int_{t_0}^{t}\hat{H}_E(t_1)dt_1}{i \hbar}}\hat{\rho}_E(t_0)e^{\frac{-\int_{t_0}^{t}\hat{H}_E(t_1)dt_1}{i \hbar}},
     \hat{\rho}_P(t)=e^{\frac{\int_{t_0}^{t}\hat{H}_P(t_1)dt_1}{i \hbar}}\hat{\rho}_P(t_0)e^{\frac{-\int_{t_0}^{t}\hat{H}_P(t_1)dt_1}{i \hbar}}, 
     \nonumber \\
\end{eqnarray}
where we have introduced antidiagonal matrix terms with the notation "adiag[,]", which start from the left bottom of the matrix, and diagonal matrix elements with the notation "diag[,]" staring from the left top of the matrix. 
We can confirm the following elementary steps as given below. 
\begin{eqnarray}
\hat{H}_E \rightarrow \hat{H}_P = \hat{R} \hat{H}_E \hat{R}^{-1}=
\hat{R} [ E_1 \ket{E_1}\bra{E_1} + E_2 \ket{E_2}\bra{E_2}+ E_{12} \ket{E_2}\bra{E_1} + E_{21} \ket{E_1}\bra{E_2} ] \hat{R}^{-1}=
 \nonumber \\
 =
  \begin{bmatrix}
        +\cos(\Theta_s) & +\sin(\Theta_s) \\
        +\sin(\Theta_s) & -\cos(\Theta_s)
    \end{bmatrix} \Bigg[ E_1 \ket{E_1}\bra{E_1} + E_2 \ket{E_2}\bra{E_2}+ E_{12} \ket{E_2}\bra{E_1} + E_{21} \ket{E_1}\bra{E_2} \Bigg]
     \begin{bmatrix}
        +\cos(\Theta_s) & - \sin(\Theta_s) \\
        +\sin(\Theta_s) & +\cos(\Theta_s)
    \end{bmatrix}
    =
 \nonumber \\
 =[ E_{P1} \ket{w_{L}}\bra{w_L} + E_{P2} \ket{w_{R}}\bra{w_{R}}+ t_{s}e^{+i \Theta_P} \ket{w_L}\bra{w_R} + t_{s}e^{-i \Theta_P} \ket{w_R}\bra{w_L} ] 
, \nonumber \\
\hat{H}_P \rightarrow \hat{H}_E = \hat{R}^{-1} \hat{H}_P \hat{R} =
 \nonumber \\
 =
 \begin{bmatrix}
        +\cos(\Theta_s) & - \sin(\Theta_s) \\
        +\sin(\Theta_s) & +\cos(\Theta_s)
    \end{bmatrix}
 \Bigg[
 E_{P1} \ket{w_{L}}\bra{w_L} + E_{P2} \ket{w_{R}}\bra{w_{R}}+ t_{s}e^{+i \Theta_P} \ket{w_L}\bra{w_R} + t_{s}e^{-i \Theta_P}
 \Bigg]
 \begin{bmatrix}
        + \cos(\Theta_s) &  + \sin(\Theta_s) \\
        - \sin(\Theta_s) &  + \cos(\Theta_s)
    \end{bmatrix} = \nonumber \\
    =
\begin{bmatrix}
        + \cos(\Theta_s) &  - \sin(\Theta_s) \\
        + \sin(\Theta_s) &  + \cos(\Theta_s)
    \end{bmatrix}
\begin{bmatrix}
        E_{P1} & t_s e^{+i\Theta_P} \\
        t_s e^{-i\Theta_P} &  + E_{P2}
\end{bmatrix}
    \begin{bmatrix}
        + \cos(\Theta_s) &  + \sin(\Theta_s) \\
        - \sin(\Theta_s) &  + \cos(\Theta_s)
    \end{bmatrix}=
    \begin{bmatrix}
        E_{1}=E_{11} & E_{21} \\
        E_{12} &  E_{2}=E_{22}
    \end{bmatrix}, \nonumber \\
    \ket{\psi}_E=
    \begin{bmatrix}
        \sqrt{p_{E1}(t)}e^{i \gamma_1(t)} & 0 \\
        0 & \sqrt{p_{E2}(t)}e^{i \gamma_2(t)}
    \end{bmatrix}
    \begin{bmatrix}
      \psi_{E1}(x) \\
      \psi_{E2}(x)
    \end{bmatrix}
    =
    \begin{bmatrix}
      \sqrt{p_{E1}(t)}e^{i \gamma_1(t)} \psi_{E1}(x) \\
      \sqrt{p_{E2}(t)}e^{i \gamma_2(t)} \psi_{E2}(x)
    \end{bmatrix} = \ket{\psi}_{E}
    \rightarrow   \ket{\psi}_{P}, \nonumber \\
    \ket{\psi}_{E}
     \rightarrow
    \ket{\psi}_{P}=\hat{R} \ket{\psi}_{E} 
    =
    \hat{R}
    \begin{bmatrix}
        \sqrt{p_{E1}(t)}e^{i \gamma_1(t)} & 0 \\
        0 & \sqrt{p_{E2}(t)}e^{i \gamma_2(t)}
    \end{bmatrix} 
    \begin{bmatrix}
      \psi_{E1}(x) \\
      \psi_{E2}(x)
    \end{bmatrix} =
    \Bigg[
     \hat{R}
    \begin{bmatrix}
        \sqrt{p_{E1}}e^{i \gamma_1} & 0 \\
        0 & \sqrt{p_{E2}}e^{i \gamma_2}
    \end{bmatrix} 
     \hat{R}^{-1} \Bigg]
    \begin{bmatrix}
      w_{L}(x) \\
      w_{R}(x)
    \end{bmatrix}, \nonumber \\
     \ket{\psi}_{P}
     \rightarrow
    \ket{\psi}_{E}=\hat{R}^{-1} \ket{\psi}_{P} 
    =
    \begin{bmatrix}
        \sqrt{p_{E1}(t)}e^{i \gamma_1(t)} & 0 \\
        0 & \sqrt{p_{E2}(t)}e^{i \gamma_2(t)}
    \end{bmatrix} 
    \begin{bmatrix}
      \psi_{E1}(x) \\
      \psi_{E2}(x)
    \end{bmatrix} =
    \Bigg[
    \begin{bmatrix}
        \sqrt{p_{E1}}e^{i \gamma_1} & 0 \\
        0 & \sqrt{p_{E2}}e^{i \gamma_2}
    \end{bmatrix} 
     \hat{R} \Bigg]
    \begin{bmatrix}
      w_{L}(x) \\
      w_{R}(x)
    \end{bmatrix}, \nonumber \\
    \Bigg[
   \ket{\psi}_E \rightarrow \ket{\psi}_P, 
   \begin{bmatrix}
       \psi_{E1}(x) \\
       \psi_{E2}(x)
    \end{bmatrix}
    \rightarrow 
     \begin{bmatrix}
       w_L(x) \\
       w_R(x)
    \end{bmatrix}
    \Bigg] :
    \begin{bmatrix}
        \sqrt{p_{E1}(t)}e^{i \gamma_1(t)} & 0 \\
        0 & \sqrt{p_{E2}(t)}e^{i \gamma_2(t)}
    \end{bmatrix} \rightarrow 
       \Bigg[
     \hat{R}
    \begin{bmatrix}
        \sqrt{p_{E1}}e^{i \gamma_1} & 0 \\
        0 & \sqrt{p_{E2}}e^{i \gamma_2}
    \end{bmatrix} 
     \hat{R}^{-1} \Bigg], \nonumber \\
     ( \hat{\rho}_E=\ket{\psi_E}\bra{\psi_E} )\rightarrow 
     ( \hat{\rho}_P= \ket{\psi_P}\bra{\psi_P} 
     =
     \hat{R}\ket{\psi_E}\bra{\psi_E}\hat{R}^{-1} =
     \hat{R}\ket{\psi_E}\bra{\psi_E}\hat{R}^{T} = \hat{R} \hat{\rho}_E \hat{R}^{T} = \hat{R} \hat{\rho}_E \hat{R}^{-1}
     )  ,  \nonumber \\
\end{eqnarray}

We denote $E_{P1}$ and $E_{P2}$ in \eqref{eq:Hp} from Hamiltonian $\hat{H}_P$ as localized energies of the left q-dot $1$ and right q-dot $2$, while antidiagonal terms are hopping energies that express single-electron movement between adjacent quantum dots. The position-based $\hat{H}_P$ matrix can be mapped into quaternion $\vb*{\hat{q}}$:
\begin{equation}
    \label{eq:Quat}
    \vb*{\hat{q}} = \begin{bmatrix}
        a + \text{i}b & c+\text{i}d \\
        -c+\text{i}d & a-\text{i}b
    \end{bmatrix} = a \hat{\sigma}_0 + i d \hat{\sigma}_1  + i c \hat{\sigma}_2 
    + i b \hat{\sigma}_3,    .
\end{equation}
what implies that quaternion is the linear combination
of 4 real-value numbers and 4 Pauli matrices, where $i=\sqrt{-1}$ is imaginary unit. 
As can easily be seen, the two matrices differ in the signs between the real and imaginary parts on the anti-diagonals, which brings to mind the multiplication of one of these matrices by the imaginary number~$i$.If we assume that by and we multiply the quaternion, we obtain the equation:
We observe that the following occurs:
\begin{equation}
    \label{eq:Rel1}
    \hat{H}_{P} = \text{i} \vb*{\hat{q}_1},
\end{equation}
%when the following conditions take place:
\begin{equation}
    \label{eq:UkladRownan}
    \begin{cases}
        a = 0,\\
        b = -E_{P1}=E_{P2}, \\
        d = t_s\cos{\theta_P},\\
        c = -t_s\sin{\theta_P}.
    \end{cases}
\end{equation}
From the equations above \eqref{eq:UkladRownan}, we conclude that the energies $E_{P1}$ and $E_{P2}$ must be equal in absolute value but opposite in sign. Thus, we will replace equation \eqref{eq:Hp} by the following expression:
\begin{equation}
    \label{eq:HpSUM}
    \hat{H}_P = \begin{bmatrix}
        - \Delta E_{P} & t_s\cos{\theta_P}+\text{i}t_s\sin{\theta_P} \\
        t_s\cos{\theta_P}-\text{i}t_s\sin{\theta_P} & \Delta E_{P}
    \end{bmatrix} + \begin{bmatrix}
        \langle E_P \rangle & 0 \\
        0 & \langle E_P \rangle
    \end{bmatrix},
\end{equation}
where $\langle E_P \rangle$~--~is~arithmetic average of $E_{P1}$ and $E_{P2}$; $\Delta E_P$~--~is~difference between $\langle E_P \rangle$ and one of the values $E_{P1}$, $E_{P2}$.
For the sake of further consideration, let us assume that $E_{P1} < E_{P2}$ and for $\Delta E_P$ we choose the smaller one, so that the expression $\Delta E_P$ is greater than 0.
Since our idea is to multiply the equation \eqref{eq:HpSUM} by $\frac{1}{\text{i}}$ and thus obtain the quaternion we note that $\langle E_P \rangle$ must be equal to 0. Fortunately for us, we are dealing with potential energy, so we can take the energy of $\langle E_P \rangle = 0$ as a reference point with impunity.
Based on equations \eqref{eq:Quat} and \eqref{eq:UkladRownan} we know that $\langle E_P \rangle = 0$ must be satisfied.
For clarity in further calculations, let us assume here that we will denote the values of $E_{P1}$ and $E_{P2}$ for which the arithmetic mean is zero by $E_{PZ1}$ and $E_{PZ2}$ respectively.
For clarity in further calculations, assume that $E_{P1}=E_{PZ1}$, $E_{P2} = E_{PZ2}$, and $\langle E_{P} \rangle = \langle E_{PZ} \rangle$ when $\langle E_P \rangle = 0$.
We will then denote the average by $\langle E_{PZ} \rangle$.
Let us therefore rewrite the system of equations \eqref{eq:UkladRownan} so that it is consistent with the equation \eqref{eq:HpSUM}. We then obtain:
Then the system of equations \eqref{eq:UkladRownan} will take the form:
\begin{equation}
    \label{eq:UkladRownan2}
    \begin{cases}
        \langle E_{PZ} \rangle = \frac{E_{PZ1}+E_{PZ2}}{2} =0,\\
        a = 0,\\
        b = \Delta E_P = -E_{PZ1} = E_{PZ2},\\
        c = t_s\sin{\theta_P},\\
        d = -t_s\cos{\theta_P}.
    \end{cases}
\end{equation}
Based on the system of equations \eqref{eq:UkladRownan2}, we can write $\vb*{\hat{q}_0}$ quaternion explicitly as:
\begin{equation}
    \label{eq:QuatEXP}
    \vb*{\hat{q}_0} = \begin{bmatrix}
        \text{i} E_{PZ2} & t_s\sin{\theta_P}-\text{i}t_s\cos{\theta_P}  \\
        -t_s\sin{\theta_P}-\text{i}t_s\cos{\theta_P} & -\text{i} E_{PZ2}      
    \end{bmatrix}.
\end{equation}
For the following steps, let us introduce the following matrix:
\begin{equation}
    \label{eq:Av}
    \hat{A}=\begin{bmatrix}
    \frac{E_{P1}+E_{P2}}{2} & 0 \\
    0 & \frac{E_{P1}+E_{P2}}{2}
    \end{bmatrix}
    =\frac{E_{P1}+E_{P2}}{2}
    \begin{bmatrix}
    1 & 0 \\
    0 & 1 
    \end{bmatrix},
\end{equation}
and such that $\hat{A}+\text{i}\vb*{\hat{q}_0}=\hat{H}_d+\hat{H}_{nd}=\hat{H}$.
We notice that Hamiltonian can be written as sum of a diagonal part with equal elements on the diagonal $\hat{H}_{d}$ plus the other remaining matrix, so $\hat{H}_{nd}$.
In such a way quantum system evolution can be characterized by following evolution operator:
\begin{multline}
\hat{U}(\tau_0,\tau)=e^{\int_{\tau_0}^{\tau}d\tau_1 \frac{1}{\text{i}\hbar}\hat{H}(\tau_1)}
=e^{\int_{\tau_0}^{\tau}d\tau_1\frac{1}{\text{i}\hbar}\hat{H}_{nd}(\tau_1)}
e^{\int_{\tau_0}^{\tau}d\tau_1\frac{1}{\text{i}\hbar}\hat{H}_{d}(\tau_1)}=e^{\frac{1}{\hbar}\int_{\tau_0}^{\tau}d\tau_1 \vb*{\hat{q}_0(\tau_1)}}e^{\frac{1}{\text{i}\hbar}\int_{\tau_0}^{\tau}d\tau_1 \hat{A}(\tau_1)}=\\=
e^{\frac{1}{\hbar}\int_{\tau_0}^{\tau}d\tau_1 \vb*{\hat{q}_0(\tau_1)}}
e^{ \frac{1}{i\hbar} \int_{\tau_0}^{\tau} d\tau_1  \frac{1}{2} (E_{P1}(\tau_1)+E_{P2}(\tau_1)) }  \text{diag}(1,1)=
e^{\frac{1}{\hbar}\int_{\tau_0}^{\tau}d\tau_1 \vb*{\hat{q}_0(\tau_1)}}e^{\frac{1}{\text{i}\hbar}\int_{\tau_0}^{\tau}d\tau_1\frac{1}{2}(E_{P1}(\tau_1)+E_{P2}(\tau_1))},
\end{multline}
where $\tau$ represents time. We have used Hausdorff-Baker formula with observation that commutator $[\hat{H}_{d},\hat{H}_{nd}]=\hat{0}$.
Finally we have the evolution of quantum state in the form:
\begin{multline}
%\begin{multline}
|\psi(\tau) \rangle_E =
\hat{U}(\tau_0,\tau)
|\psi(\tau_0) \rangle_E
=e^{\frac{1}{\hbar}\int_{\tau_0}^{\tau}d\tau_1 \vb*{\hat{q}_0(\tau_1)}}e^{\frac{1}{\text{i}\hbar}\int_{\tau_0}^{\tau}d\tau_1\frac{1}{2}(E_{P1}(\tau_1)+E_{P2}(\tau_1))}|\psi(\tau_0)\rangle_E =\\=e^{\frac{1}{\hbar}\int_{\tau_0}^{\tau}d\tau_1 \vb*{\hat{q}_0(\tau_1)}} \begin{bmatrix}
    \int_{\tau_0}^{\tau}d\tau_1\frac{1}{2}(E_{P1}(\tau_1)+E_{P2}(\tau_1)) & 0\\ 0& \int_{\tau_0}^{\tau}d\tau_1\frac{1}{2}(E_{P1}(\tau_1)+E_{P2}(\tau_1))
\end{bmatrix}\begin{bmatrix}
    \Psi_1(\tau_0)=\sqrt{p_{E1}(\tau_0)}e^{i \gamma_1(\tau_0)}\psi_{E1}(x)
    \\\Psi_2(\tau_0)=\sqrt{p_{E2}(\tau_0)}e^{i \gamma_2(\tau_0)}\psi_{E2}(x)
\end{bmatrix} .
=\\=e^{\int_{t0}^{t}d\tau\frac{1}{i\hbar}\frac{E_{P1}(\tau)+E_{P2}(\tau)}{2}}[e^{\int_{t0}^{t}\frac{1}{\hbar }d\tau \hat{q}_1(\tau)\hat{q}_2(\tau)}]e^{\frac{1}{i \hbar}}|\psi(t_0)\rangle
\end{multline}
%\end{multline}
We can identify the energy density matrix $\hat{\rho}_E = |\psi(\tau) \rangle_E  \langle \psi(\tau)|_E $ as 
\begin{multline} \label{eq:0}
 \hat{\rho}_E =
\Big[ |\psi(\tau) \rangle  \langle \psi(\tau)| \Big] _E =
\hat{U}(\tau_0,\tau)
[ |\psi(\tau_0) \rangle 
\langle \psi(\tau_0) \rangle ]_E \hat{U}^{\dag}(\tau_0,\tau)
= \nonumber \\
=e^{\frac{1}{\hbar}\int_{\tau_0}^{\tau}d\tau_1 \vb*{\hat{q}_0(\tau_1)}}e^{\frac{1}{\text{i}\hbar}\int_{\tau_0}^{\tau}d\tau_1\frac{1}{2}(E_{P1}(\tau_1)+E_{P2}(\tau_1))} |\psi(\tau_0)\rangle_E \langle \psi(\tau)|_E 
e^{\frac{1}{\hbar}\int_{\tau_0}^{\tau}d\tau_3 \vb*{\hat{q}_0^{*}(\tau_3)}}e^{-\frac{1}{\text{i}\hbar}\int_{\tau_0}^{\tau}d\tau_2\frac{1}{2}(E_{P1}(\tau_2)+E_{P2}(\tau_2))} 
= \nonumber \\
=e^{\frac{1}{\hbar}\int_{\tau_0}^{\tau}d\tau_1 \vb*{\hat{q}_0(\tau_1)}} |\psi(\tau_0)\rangle_E \langle \psi(\tau)|_E 
e^{\frac{1}{\hbar}\int_{\tau_0}^{\tau}d\tau_3 \vb*{\hat{q}_0^{*}(\tau_3)}} 
= \nonumber \\
%%\\
=e^{\frac{1}{\hbar}\int_{\tau_0}^{\tau}d\tau_1 \vb*{\hat{q}_0(\tau_1)}} 
\begin{bmatrix}
    \sqrt{p_{E1}}e^{i \gamma_1(\tau_0)}\psi_{E1}(x)
    \\
    \sqrt{p_{E2}}e^{i \gamma_2(\tau_0)}\psi_{E2}(x)
\end{bmatrix} 
%%%\nonumber \\
\begin{bmatrix}
    \sqrt{p_{E1}}e^{-i \gamma_1(\tau_0)}\psi_{E1}^{*}(x) & \sqrt{p_{E2}}e^{-i \gamma_2(\tau_0)}\psi_{E2}^{*}(x) \\
\end{bmatrix} 
e^{\frac{1}{\hbar}\int_{\tau_0}^{\tau}d\tau_1 \vb*{\hat{q}_0^{*}(\tau_1)}} = \nonumber \\
=
e^{\frac{1}{\hbar}\int_{\tau_0}^{\tau}d\tau_1 \vb*{\hat{q}_0(\tau_1)}}
\begin{bmatrix}
    p_{E1}|\psi_{E1}(x)|^2 & \sqrt{p_{E1}}\sqrt{p_{E2}}e^{+i[ \gamma_1(\tau_0) - \gamma_2(\tau_0) ]}\psi_{E1}(x)\psi_{E2}^{*}(x)   \\
    \\
    \sqrt{p_{E1}}\sqrt{p_{E2}}e^{-i[ \gamma_1(\tau_0) - \gamma_2(\tau_0) ]}\psi_{E1}^{*}(x)\psi_{E2}(x) 
     & p_{E2}|\psi_{E2}(x)|^2
\end{bmatrix} 
e^{\frac{1}{\hbar}\int_{\tau_0}^{\tau}d\tau_1 \vb*{\hat{q}_0^{*}(\tau_1)}}
.
\end{multline}
Due to the fact that equation of motion of 
matrix density can be quaternized we have 
the following equivalence
\begin{eqnarray} \label{eq:1}
i\hbar \frac{d}{dt}\hat{\rho}= [\hat{H},\hat{\rho}], 
-i\hbar \frac{d}{dt}\hat{\rho}= -[\hat{H},\hat{\rho}], -i\hbar \frac{d}{dt}\hat{\rho}= [i \hat{H},\hat{\rho} i ], -\hbar \frac{d}{dt}(\hat{\rho}i)= [ (i \hat{H}), (\hat{\rho} i)]=-[(\hat{\rho} i),(i \hat{H})], 
\end{eqnarray}
and final conclusion points the equation of motion 
of quaternized density matrix $\hat{\rho}_{QT}=i\hat{\rho} $ in the real time as in contrary to Schroedinger equation that is diffusion equation in imaginary time. Therefore, the key observation is that 
\begin{eqnarray} \label{eq:2}
\hbar \frac{d}{dt}\hat{\rho}_{QT}=[\hat{\rho}_{QT},(i \hat{H})]=-[(i \hat{H}),\hat{\rho}_{QT}]=[(\frac{1}{i} \hat{H}),\hat{\rho}_{QT}], %%%%\hat{\rho}_{QT}(t)=e^{}\hat{\rho}_{QT}(t_0), 
\nonumber \\
\hat{\rho}_{E,QT}(\tau)=e^{+\frac{1}{\hbar}\int_{\tau_0}^{\tau}d\tau_1 \vb*{\hat{q}_0(\tau_1)}}\hat{\rho}_{QT}(\tau_0)e^{-\frac{1}{\hbar}\int_{\tau_0}^{\tau}d\tau_1 \vb*{\hat{q}_0^{*}(\tau_1)}}=
e^{\frac{1}{\hbar i}\int_{\tau_0}^{\tau}d\tau_1 \vb*{\hat{H}(\tau_1)}}\hat{\rho}_{QT}(\tau_0)e^{-\frac{1}{\hbar i}\int_{\tau_0}^{\tau}d\tau_1 \vb*{\hat{H}_0^{*}(\tau_1)}}.
\end{eqnarray}
Fundamental decomposition of quaternionic density matrix of eigenergy qubit is as following
\begin{eqnarray} \label{eq:3}
\hat{\rho}_{QT}(\tau_0)=i \hat{\rho}_E(\tau_0)=i\hat{\sigma}_0 \frac{ p_{E2}|\psi_{E2}(x)|^2 + p_{E1}|\psi_{E1}(x)|^2}{2}-i \frac{ p_{E2}|\psi_{E2}(x)|^2 - p_{E1}|\psi_{E1}(x)|^2}{2}\hat{\sigma}_3+ \nonumber \\
-i \hat{\sigma_1} \sqrt{p_{E1}}\sqrt{p_{E2}} Im[e^{-i(\gamma_2(\tau_0) - \gamma_1(\tau_0))} \psi_{E1}(x)\psi_{E2}^{*}(x) ] + i \sigma_2 \sqrt{p_{E1}}\sqrt{p_{E2}} Re[e^{-i(\gamma_2(\tau_0) - \gamma_1(\tau_0))} \psi_{E1}(x)\psi_{E2}^{*}(x) ],
\end{eqnarray}
, while in the case of position-based qubit is given as
\begin{eqnarray} \label{eq:4}
\hat{\rho}_{P,QT}(\tau_0)=i\hat{R} \hat{\rho}_E(\tau_0) \hat{R}^{-1}=i  \hat{R}\hat{\sigma}_0 \hat{R}^{-1} \frac{ p_{E2}|\psi_{E2}(x)|^2 + p_{E1}|\psi_{E1}(x)|^2}{2}-i \frac{ p_{E2}|\psi_{E2}(x)|^2 - p_{E1}|\psi_{E1}(x)|^2}{2}\hat{\sigma}_3+ \nonumber \\
-i  \hat{R}\hat{\sigma_1} \hat{R}^{-1} \sqrt{p_{E1}}\sqrt{p_{E2}} Im[e^{-i(\gamma_2(\tau_0) - \gamma_1(\tau_0))} \psi_{E1}(x)\psi_{E2}^{*}(x) ] + i  \hat{R} \sigma_2 \hat{R}^{-1} \sqrt{p_{E1}}\sqrt{p_{E2}} Re[e^{-i(\gamma_2(\tau_0) - \gamma_1(\tau_0))} \psi_{E1}(x)\psi_{E2}^{*}(x) ].
\end{eqnarray}
\section{Case of many body systems in quaternionic representation}
Let us define a tensor product of k Hilbert spaces expressing system of k qubits. It is corresponding to formally known a generalized double Kronecker sum and given by the following expression: 
\begin{multline}
    \chi_{o,p}^{r,s}\left(\hat{M}_n,\hat{M}_m,E(n,m) \right) \coloneq \sum_{n=o}^r\sum_{m=p}^s \left( \hat{I}_1 \otimes_K \hat{I}_2 \otimes_K \ldots \otimes_K \hat{I}_{n-1} \otimes_K E(n,m) \cdot \hat{M}_{n} \otimes_K \hat{I}_{n+1} \otimes_K \ldots \right. \\ \left. \ldots \otimes_K \hat{I}_{m-1} \otimes_K \hat{M}_{m} \otimes_K \hat{I}_{m+1} \otimes_K \ldots \otimes_K \hat{I}_t \right)
\end{multline}
where: $t$~--~greater value with $r$ and $s$; $\hat{I}_\alpha$~--~two-by-two unit matrix; $E(n,m)$~--~function which assigns to any pair of numbers $n,m \in \mathbb{N}^+$ a given value; $\hat{M_n}, \hat{M}_m$~--~fixed type matrices with values dependent on the values of the $n,m\in\mathbb{N}^+$.
When the summation produces an expression with equal subscripts $n$ and $m$, the expression will be written as:
\begin{equation}
    \chi_{o,p}^{r,s}\left( \hat{M}_n,\hat{M}_m,E(n,m) \right) = \ldots +\hat{I}_1 \otimes_K \hat{I}_2 \otimes_K \ldots \otimes_K \hat{I}_{l-1} \otimes_K E(l,l) \cdot \hat{M}_{n=l} \cdot \hat{M}_{m=l} \otimes_K \hat{I}_{l+1} \otimes_K \ldots \otimes_K \hat{I}_t +\ldots
\end{equation}
\subsection{Case of non-interacting k-qubits system in quaternion representation}
The Hamiltonian $\hat{H}$ of system of $k$ non-interacting quantum bodies (as qubits) denotes to:
\begin{equation}
    \hat{H} = \chi_{1,k}^{k,k}\left( \hat{H}_n,\hat{I}_m,1 \right),
\end{equation}
where $\hat{H}_n$ is a Hamiltonian of the $n$-th dot. By the analogy to \eqref{eq:Rel1}, we construct a definition for $k$ bodies as:
\begin{equation}
\label{eq:HkB}
    \hat{H} = \chi_{1,k}^{k,k}\left( \langle E_n \rangle \cdot \hat{\sigma}_0 + \text{i} \vb*{\hat{q}_n} ,\hat{I}_m,1 \right)  = \chi_{1,k}^{k,k}\left( \hat{\sigma}_0, \hat{I}_m, \langle E_n \rangle \right) + \chi_{1,k}^{k,k}\left( \vb*{\hat{q}_n}, \hat{I}_m, \text{i} \right).
\end{equation}
In such a way quantum system evolution can be described by the following operator:
\begin{multline}
    \exp\left({\int_{t0}^{t}d\tau \frac{1}{\text{i}\hbar}\hat{H}(\tau)}\right) = \exp\left({\int_{t0}^{t}d\tau\frac{1}{\text{i}\hbar}\chi_{1,k}^{k,k}\left( \vb*{\hat{q}_n}, \hat{I}_m, \text{i} \right)}\right) \cdot \exp\left({\int_{t0}^{t}d\tau\frac{1}{\text{i}\hbar}\chi_{1,k}^{k,k}\left( \hat{\sigma}_0, \hat{I}_m, \langle E_n \rangle \right)}\right)=\\=\exp\left({\frac{1}{\hbar}\int_{t0}^{t}d\tau \chi_{1,k}^{k,k}\left( \vb*{\hat{q}_n}, \hat{I}_m, 1 \right)}\right) \cdot \exp\left({\frac{1}{\text{i}\hbar}\int_{t0}^{t}d\tau \left( \sum_{n=1}^k \langle E_n(\tau) \rangle \right) \cdot \chi_{1,k}^{k,k}\left( \hat{\sigma}_0,\hat{I}_m,1 \right)}\right) =\\=
\exp\left({\frac{1}{\hbar}\int_{t0}^{t}d\tau \chi_{1,k}^{k,k}\left( \vb*{\hat{q}_n}, \hat{I}_m, 1 \right)}\right) \cdot \exp\left({\frac{1}{\text{i}\hbar}\int_{t0}^{t}d\tau \left( \sum_{n=1}^k \langle E_n(\tau) \rangle \right) \text{diag}(1,1,\ldots,1)}\right)=\\=
\exp\left({\frac{1}{\hbar}\int_{t0}^{t}d\tau \chi_{1,k}^{k,k}\left( \vb*{\hat{q}_n}, \hat{I}_m, 1 \right)}\right) \cdot \exp\left({\frac{1}{\text{i}\hbar}\int_{t0}^{t}d\tau\left( \sum_{n=1}^k \langle E_n(\tau) \rangle \right)}\right).
\end{multline}
Finally we have the evolution of quantum state in the form:
\begin{equation}
    |\psi(\tau)\rangle=\exp\left({\frac{1}{\hbar}\int_{t0}^{t}d\tau \chi_{1,k}^{k,k}\left( \vb*{\hat{q}_n}, \hat{I}_m, 1 \right)}\right) \cdot \exp\left({\frac{1}{\text{i}\hbar}\int_{t0}^{t}d\tau\left( \sum_{n=1}^k \langle E_n(\tau) \rangle \right)}\right)|\psi(\tau_0)\rangle.
\end{equation}
\subsection{Electrostatic interaction between two position-based qubits }
We consider two A and B interacting position-based qubits made from double adjacent quantum dots (correspondingly denoted by pairs of adjacent dots  $(x_{A,L},x_{A,R} )$ and $(x_{B,L},x_{B,R} )$) and we 
can encounter the following effective position-based Hamiltonian given as 
\begin{eqnarray}
\hat{H}_P = \hat{H}_{PA} + \hat{H}_{PB} + \hat{H}_{PA-PB} = \nonumber \\
=
[ E_{PA,L}\ket{x_{A,L}}\bra{x_{A,L}}  + E_{PA,R}\ket{x_{A,R}}\bra{x_{A,R}} +t_{sA}e^{i \Theta_{P,A}}\ket{x_{A,L}}\bra{x_{A,R}} + t_{sA}e^{-i \Theta_{P,A}}\ket{x_{A,R}}\bra{x_{A,L}}  ] \otimes_K \hat{I}_B +  \nonumber \\
+ \hat{I}_A \times
[ E_{PB,L}\ket{x_{B,L}}\bra{x_{B,L}}  + E_{PB,R}\ket{x_{B,R}}\bra{x_{B,R}} +t_{sB}e^{i \Theta_{P,B}}\ket{x_{B,L}}\bra{x_{B,R}} + t_{sB}e^{-i \Theta_{P,B}}\ket{x_{B,R}}\bra{x_{B,L}}  ] + \nonumber \\
+ E_C(x_{A,L},x_{B,L}) [ \ket{x_{A,L}}\bra{x_{A,L}} ] \otimes_K [ \ket{x_{B,L}}\bra{x_{B,L}} ] +  E_C(x_{A,L},x_{B,R}) [ \ket{x_{A,L}}\bra{x_{A,L}} ]  \otimes_K [ \ket{x_{B,R}}\bra{x_{B,R}} ] + \nonumber \\
+ E_C(x_{A,R},x_{B,L}) [ \ket{x_{A,R}}\bra{x_{A,R}} ] \otimes_K [ \ket{x_{B,L}}\bra{x_{B,L}} ] +  E_C(x_{A,R},x_{B,R}) [ \ket{x_{A,R}}\bra{x_{A,R}} ]  \otimes_K [ \ket{x_{B,R}}\bra{x_{B,R}} ] = \nonumber \\
=
  \begin{bmatrix}
    E_{PA,L} + E_{PB,L}  & t_{sB}e^{+i\Theta_{P,B}} & t_{sA}e^{+i\Theta_{P,A}} & 0 \\
    t_{sB}e^{-i\Theta_{P,B}} & E_{PA,L} + E_{PB,R}  & 0 & t_{sA}e^{+i\Theta_{P,A}} \\
    t_{sA}e^{-i\Theta_{P,A}} & 0 & E_{PA,L} + E_{PB,L}  & t_{sB}e^{+i\Theta_{P,B}} \\
    0 & t_{sA}e^{-i\Theta_{P,A}} & t_{sB}e^{-i\Theta_{P,B}} & E_{PA,R} + E_{PB,R}  \\
    \end{bmatrix}
    + \nonumber \\
     diag ( E_C(x_{A,L},x_{B,L}) , E_C(x_{A,L},x_{B,R}) , E_C(x_{A,R},x_{B,L}) , E_C(x_{A,R},x_{B,R}) ) =
     \nonumber \\
       \begin{bmatrix}
    Q_1  & h_2 & h_1 & 0 \\
    h_2^{*} & Q_2  & 0 & h_1 \\
    h_1^{*} & 0 & Q_3  & h_2 \\
    0 & h_1^{*} & h_2^{*} & Q_4  \\
    \end{bmatrix} , \nonumber \\
    Q_1=  E_{PA,L} + E_{PB,L} + E_C(x_{A,L},x_{B,L}),  E_C(x_{A,L},x_{B,L})=\frac{e^2}{d(x_{A,L},x_{B,L})}, \nonumber \\
    Q_2=  E_{PA,L} + E_{PB,R} + E_C(x_{A,L},x_{B,R}), E_C(x_{A,L},x_{B,R})=\frac{e^2}{d(x_{A,L},x_{B,R})}, \nonumber \\
    Q_3=  E_{PA,R} + E_{PB,L} + E_C(x_{A,R},x_{B,L}), E_C(x_{A,R},x_{B,L})=\frac{e^2}{d(x_{A,R},x_{B,L})}, \nonumber \\
    Q_4=  E_{PA,R} + E_{PB,R} + E_C(x_{A,R},x_{B,R}), E_C(x_{A,R},x_{B,R})=\frac{e^2}{d(x_{A,R},x_{B,R})}, \nonumber \\
   h_1 = t_{sA}e^{+i\Theta_{P,A}} , \nonumber \\
   h_2 = t_{sB}e^{+i\Theta_{P,B}} , \nonumber \\
\end{eqnarray}
where ( $E_C(x_{A,L},x_{B,L})$,  $E_C(x_{A,L},x_{B,L})$ ,  $E_C(x_{A,L},x_{B,L})$ ,  $E_C(x_{A,L},x_{B,L})$) are Coulomb interaction
terms between maximum localized Wanniers (W) function of corresponding qubits (left W-function of qubit A with left W-function of qubit B), 
(left W-function of qubit A with right W-function of qubit B), (right W-function of qubit A with left W-function of qubit B),
(right W-function of qubit A with right W-function of qubit B). In the oversimplified case of two symmetric qubits 
all localized energies are the same $E_p=E_{PA,L}=E_{PA,R}=E_{PB,L}=E_{PB,R}$ and also $h_1=h_2=h=h^{*}$ and also for parallel qubits A and B we encounter $E_{C1}= E_C(x_{A,L},x_{B,L})=E_C(x_{A,R},x_{B,R})$ and $E_{C2}= E_C(x_{A,L},x_{B,R})=E_C(x_{A,L},x_{B,R})$. This ends up in 
more simplified version of the position-based Hamiltonian matrix for 2 coupled qubits A and B in the form as : 
\begin{eqnarray}
   \begin{bmatrix}
    Q_5  & h & h & 0 \\
    h & Q_6  & 0 & h \\
    h & 0 & Q_6  & h \\
    0 & h & h & Q_5  \\
    \end{bmatrix} 
    = \frac{Q_5+Q_6}{2}
      \begin{bmatrix}
    1  & 0 & 0 & 0 \\
    0 & 1 & 0 & 0 \\
    0 & 0 & 1  & 0 \\
    0 & 0 & 0 & 1 \\
    \end{bmatrix} + \frac{1}{2}
     \begin{bmatrix}
      +(Q_5 - Q_6) & 2h & 2h & 0 \\
    2h & -(Q_5 - Q_6)  & 0 & 2h \\
    2h & 0 & +(Q_5 - Q_6)  & 2h \\
    0 & 2h & 2h & -(Q_5 - Q_6)  \\
    \end{bmatrix} = \nonumber \\
    =\frac{Q_5+Q_6}{2}
     \begin{bmatrix}
        1  & 0 \\
        0 & 1  \\
    \end{bmatrix}
     \otimes_K
     \begin{bmatrix}
        1  & 0 \\
        0 & 1  \\
    \end{bmatrix} +
    \frac{Q_5-Q_6}{2}
    \begin{bmatrix}
        1  & 0 \\
        0 & 1  \\
    \end{bmatrix}
     \otimes_K
     \begin{bmatrix}
        1  & 0 \\
        0 & -1  \\
    \end{bmatrix} + h
     \begin{bmatrix}
        1  & 0 \\
        0 & 1  \\
    \end{bmatrix}
     \otimes_K
     \begin{bmatrix}
        0  & 1 \\
        1  & 0  \\
    \end{bmatrix}
    +h
    \begin{bmatrix}
        0  & 1 \\
        1  & 0  \\
    \end{bmatrix}
     \otimes_K
      \begin{bmatrix}
        1  & 0 \\
        0  & 1  \\
    \end{bmatrix} = \nonumber \\
    = \frac{Q_5+Q_6}{2}
     \begin{bmatrix}
        1  & 0 \\
        0 & 1  \\
    \end{bmatrix}
     \otimes_K
     \begin{bmatrix}
        1  & 0 \\
        0 & 1  \\
    \end{bmatrix} -i
    \frac{Q_5-Q_6}{2}
    \begin{bmatrix}
        1  & 0 \\
        0 & 1  \\
    \end{bmatrix}
     \otimes_K
     \begin{bmatrix}
        i  & 0 \\
        0 & -i  \\
    \end{bmatrix} -i h
      \begin{bmatrix}
        1  & 0 \\
        0 & 1  \\
    \end{bmatrix}
     \otimes_K
     \begin{bmatrix}
        0  & i \\
        i  & 0  \\
    \end{bmatrix}
    -i h
    \begin{bmatrix}
        0  & i \\
        i  & 0  \\
    \end{bmatrix}
     \otimes_K
      \begin{bmatrix}
        1  & 0 \\
        0  & 1  \\
    \end{bmatrix} = \nonumber \\
    = \hat{q}_{A1}  \otimes_K \hat{q}_{B1} + i\hat{q}_{A2}  \otimes_K \hat{q}_{B2} + i\hat{q}_{A3}  \otimes_K \hat{q}_{B3} + i\hat{q}_{A4}  \otimes_K \hat{q}_{B4}, \nonumber \\
     \hat{q}_{A1} = 
     \frac{Q_5+Q_6}{2}
     \begin{bmatrix}
        1  & 0 \\
        0 & 1  \\
    \end{bmatrix}, 
 \hat{q}_{B1} = 
     \begin{bmatrix}
       1  & 0 \\
        0 & 1  \\
    \end{bmatrix},     
     \hat{q}_{A2} = 
     \frac{Q_5-Q_6}{2}
     \begin{bmatrix}
        1  & 0 \\
        0 & 1  \\
    \end{bmatrix}, 
     \hat{q}_{B2} =  
     \begin{bmatrix}
        i  & 0 \\
        0 & -i  \\
    \end{bmatrix}, 
     \hat{q}_{A3} = h  
     \begin{bmatrix}
        1  & 0 \\
        0 & 1  \\
    \end{bmatrix} ,
     \hat{q}_{B3} =   
     \begin{bmatrix}
        0  & i \\
        i & 0  \\
    \end{bmatrix} , \nonumber \\
    \hat{q}_{A4} = -h 
     \begin{bmatrix}
        0  & i \\
        i & 0  \\
    \end{bmatrix} ,
     \hat{q}_{B4} =   
     \begin{bmatrix}
        1  & 0 \\
        0 & 1  \\
    \end{bmatrix}
\end{eqnarray}
In case of 2 interacting bodies (even numbers of interacting quantum bodies as 2,4,6, .. ) we always search for quaternionic representation of time evolution operator of non-diagonal $\hat{H}_{ND}$ in the form 
\begin{eqnarray}
\exp(\frac{-i}{\hbar} \int_{t0}^{t} dt_1 \hat{H}_P(t_1)_{ND} ) = \exp(\frac{-i}{\hbar} \int_{t0}^{t} dt_1 \sum_s \hat{q}_{A,s} \otimes_K \hat{q}_{B,s} )  
\end{eqnarray}
that is different from single qubit form.  In case of single quantutm body as isolated qubit or odd number of interacting bodies (1,3,5,7 ..) we expect 
\begin{eqnarray}
 \exp(\frac{-i}{\hbar} \int_{t0}^{t} dt_1 \hat{H}_P(t_1)_{ND}  ) = \exp(\frac{-1}{\hbar} \int_{t0}^{t} dt_1 \sum_s \hat{q}_{A,s} \otimes_K \hat{q}_{B,s} )   
\end{eqnarray}
% .
%
We notice that the Kronecker product of 2 quaternions $\hat{q}_A$ and $\hat{q}_B$ can be given in the general form as 
\begin{eqnarray}
\hat{q}_A \otimes_K\hat{q}_B=
\begin{bmatrix}
        a_1 + i b_1  & i d_1 +c_1 \\
        i d_1 - c_1  & a_1 - i b_1  \\
    \end{bmatrix}
  \otimes_K
      \begin{bmatrix}
         a_2 + i b_2  & i d_2 + c_2 \\
        i d_2 - c_2  & a_2 - i b_2  \\
    \end{bmatrix} =
    \nonumber \\
    =
      \left(
\begin{array}{cccc}
 ( a_{1}+i b_{1}) (+a_{2}+i b_{2}) & (a_{1}+ib_{1}) (c_{2}+id_{2}) & (c_{1}+i
   d_{1}) (a_{2}+i b_{2}) & (c_{1}+i d_{1}) (c_{2}+i d_{2}) \\
 (a_{1}+i b_{1}) (-c_{2}+i d_{2}) & (a_{1}+i b_{1}) (a_{2}-i b_{2}) & (c_{1}+i
   d_{1}) (-c_{2}+i d_{2}) & (c_{1}+i d_{1}) (a_{2}-i b_{2}) \\
 (-c_{1}+i d_{1}) (a_{2}+i b_{2}) & (-c_{1}+i d_{1}) (i d_{2}+ c_{2}) & (a_{1}-i
   b_{1}) (a_{2}+i b_{2}) & (a_{1}-i b_{1}) (c_{2}+i d_{2}) \\
 (-c_{1}+i d_{1}) (-c_{2}+i d_{2}) &  (-c_{1}+i d_{1})(a_{2}-i b_{2})  & (a_{1}-i
   b_{1}) (-c_{2}+i d_{2}) & (a_{1}-i b_{1}) (a_{2}-i b_{2}) \\
\end{array}
\right) .
\end{eqnarray}
Consequently we have 
\begin{eqnarray}
(\hat{q}_A \otimes_K\hat{q}_B)(\hat{q}_C \otimes_K\hat{q}_D)= 
([\hat{q}_A \hat{q}_C] \otimes_K [\hat{q}_B\hat{q}_D])= \nonumber \\
=
\Bigg[
\begin{bmatrix}
        a_1 + i b_1  & i d_1 +c_1 \\
        i d_1 - c_1  & a_1 - i b_1  \\
    \end{bmatrix}
\begin{bmatrix}
        a_3 + i b_3  & i d_3 +c_3 \\
        i d_3 - c_3  & a_3 - i b_3  \\
    \end{bmatrix}    
    \Bigg]
  \otimes_K
  \Bigg[
      \begin{bmatrix}
         a_2 + i b_2  & i d_2 + c_2 \\
        i d_2 - c_2  & a_2 - i b_2  \\
    \end{bmatrix} 
    \begin{bmatrix}
        a_4 + i b_4  & i d_4 +c_4 \\
        i d_4 - c_4  & a_4 - i b_4  \\
    \end{bmatrix}.
    \Bigg]
    %\nonumber \\
    %=
\end{eqnarray}    
\subsection{Electrostatic interaction between three position-based qubits }
In case of 3 interacting bodies we have 
\begin{eqnarray}
\hat{H}_P = \hat{H}_{PA} + \hat{H}_{PB} +
 \hat{H}_{PC} +
\hat{H}_{PA-PB} +
\hat{H}_{PA-PC} +
\hat{H}_{PB-PC} +
= \nonumber \\
=
[ E_{PA,L}\ket{x_{A,L}}\bra{x_{A,L}}  + E_{PA,R}\ket{x_{A,R}}\bra{x_{A,R}} +t_{sA}e^{i \Theta_{P,A}}\ket{x_{A,L}}\bra{x_{A,R}} + t_{sA}e^{-i \Theta_{P,A}}\ket{x_{A,R}}\bra{x_{A,L}}  ] \otimes_K \hat{I}_{B} \otimes_K \hat{I}_{C} +  \nonumber \\
+%
\hat{I}_{A} \otimes_K
[ E_{PB,L}\ket{x_{B,L}}\bra{x_{B,L}}  + E_{PB,R}\ket{x_{B,R}}\bra{x_{B,R}}
%%%%+
%%%%\nonumber \\
+t_{sB}e^{i \Theta_{P,B}}\ket{x_{B,L}}\bra{x_{B,R}} + t_{sB}e^{-i \Theta_{P,B}}\ket{x_{B,R}}\bra{x_{B,L}}  ] \times \hat{I}_C   +
\nonumber \\
+
\hat{I}_{A} \otimes_K \hat{I}_{B} \otimes_K
[ E_{PC,L}\ket{x_{C,L}}\bra{x_{C,L}}  + E_{PC,R}\ket{x_{C,R}}\bra{x_{C,R}}
+
%%%%\nonumber \\
+t_{sC}e^{i \Theta_{P,C}}\ket{x_{C,L}}\bra{x_{C,R}} + t_{sC}e^{-i \Theta_{P,C}}\ket{x_{C,R}}\bra{x_{C,L}}  ]   +
 \nonumber \\
 + [ E_C(x_{AL},x_{BL}) [ \ket{x_{AL}}\bra{x_{AL}} ] \otimes_K  [\ket{x_{BL}}\bra{x_{BL}}  ]
 \otimes_K \hat{I}_{C} + E_C(x_{AL},x_{BR}) [ \ket{x_{AL}}\bra{x_{AL}} ]  \otimes_K  [\ket{x_{BR}}\bra{x_{BR}} ] \otimes_K \hat{I}_{C}  + \nonumber \\
 +  E_C(x_{AR},x_{BR}) [ \ket{x_{AR}}\bra{x_{AR}} ] \otimes_K  [\ket{x_{BR}}\bra{x_{BR}}  ]
 \otimes_K  \hat{I}_{C}
 + E_C(x_{AR},x_{BL}) [ \ket{x_{AR}}\bra{x_{AR}} ]  
 ]  [ \ket{x_{BL}}\bra{x_{BL}} ]  
 ]  
 \otimes_K  \hat{I}_{C}  
 + \nonumber \\
\hat{I}_{A}
\otimes_K 
 [ + E_C(x_{BL},x_{CL})  [ \ket{x_{BL}}\bra{x_{BL}} ] \otimes_K  [ \ket{x_{CL}}\bra{x_{CL}}  ]  %+ \nonumber \\
+ \hat{I}_{A}
\otimes_K 
 [ + E_C(x_{BL},x_{CR})  [ \ket{x_{BL}}\bra{x_{BL}} ] \otimes_K  [ \ket{x_{CR}}\bra{x_{CR}}  ] ] + \nonumber \\
+ \hat{I}_{A}
\otimes_K 
 [ + E_C(x_{BR},x_{CL})  [ \ket{x_{BR}}\bra{x_{BR}} ] \otimes_K  [ \ket{x_{CL}}\bra{x_{CL}}  ]]  % + \nonumber \\
+ \hat{I}_{A}
\otimes_K 
 [+ E_C(x_{BR},x_{CR})  [ \ket{x_{BR}}\bra{x_{BR}} ] \otimes_K  [ \ket{x_{CR}}\bra{x_{CR}}  ] ] + \nonumber \\
+
 E_C(x_{AL},x_{CL}) 
 [  \ket{x_{AL}}\bra{x_{AL}}  ] \times [  \ket{x_{BL}}\bra{x_{BL}} + \ket{x_{BR}}\bra{x_{BR}}  ] \times [  \ket{x_{CL}}\bra{x_{CL}}  ] + \nonumber \\
+
 E_C(x_{AL},x_{CR}) 
 [  \ket{x_{AL}}\bra{x_{AL}}  ] \times [  \ket{x_{BL}}\bra{x_{BL}} + \ket{x_{BR}}\bra{x_{BR}}  ] \times [  \ket{x_{CR}}\bra{x_{CR}}  ] +
 \nonumber \\
 +
E_C(x_{AR},x_{CL}) 
 [  \ket{x_{AR}}\bra{x_{AR}}  ] \times [  \ket{x_{BL}}\bra{x_{BL}} + \ket{x_{BR}}\bra{x_{BR}}  ] \times [  \ket{x_{CL}}\bra{x_{CL}}  ] + 
 \nonumber \\
   E_C(x_{AR},x_{CR}) 
 [  \ket{x_{AR}}\bra{x_{AR}}  ] \times [  \ket{x_{BL}}\bra{x_{BL}} + \ket{x_{BR}}\bra{x_{BR}}  ] \times [  \ket{x_{CR}}\bra{x_{CR}}  ]
= \nonumber \\
\begin{bmatrix}
        E_{pA,L} & t_{sA}e^{+i\Theta_{P,A}} \\
        t_{sA}e^{-i\Theta_{P,A}}  & E_{pA,R}   \\
    \end{bmatrix}
    \otimes_K
    \begin{bmatrix}
        1 & 0 \\
        0  & 1   \\
    \end{bmatrix}
     \otimes_K
    \begin{bmatrix}
        1 & 0 \\
        0  & 1   \\
    \end{bmatrix}
    +
    \begin{bmatrix}
        1 & 0 \\
        0  & 1   \\
    \end{bmatrix}
    \otimes_K
 \begin{bmatrix}
        E_{pB,L} & t_{sB}e^{+i\Theta_{P,B}} \\
        t_{sB}e^{-i\Theta_{P,B}}  & E_{pB,R}   \\
    \end{bmatrix}
    \otimes_K
     \begin{bmatrix}
        1 & 0 \\
        0  & 1   \\
    \end{bmatrix}
    + \nonumber \\
    +
        \begin{bmatrix}
        1 & 0 \\
        0  & 1   \\
    \end{bmatrix}
        \otimes_K
           \begin{bmatrix}
        1 & 0 \\
        0  & 1   \\
    \end{bmatrix}
        \otimes_K
        \begin{bmatrix}
        E_{pC,L} & t_{sC}e^{+i\Theta_{P,C}} \\
        t_{sC}e^{-i\Theta_{P,C}}  & E_{pC,R}   \\
    \end{bmatrix}
    +
      % + \nonumber \\
% \tiny
\end{eqnarray}
\small
$ + diag[ E_C(x_{AL},x_{BL}),E_C(x_{AL},x_{BL}), E_C(x_{AL},x_{BR}),E_C(x_{AL},x_{BR}),E_C(x_{AR},x_{BL}),E_C(x_{AR},x_{BL}),E_C(x_{AR},x_{BL}),E_C(x_{AR},x_{BL}) ] $
$
+diag(E_C(x_{LA},x_{CL}),E_C(x_{LA},x_{CR}),E_C(x_{LA},x_{CL}],E_C(x_{LA},x_{CR}),E_C(x_{RA},x_{CL}),E_C(x_{RA},x_{CR}),E_C(x_{RA},x_{CL}),E_C(x_{RA},x_{CR}) ]$
$ + diag [ E_C(x_{BL},x_{CL}),E_C(x_{BL},x_{CR}), E_C(x_{BR},x_{CL}),E_C(x_{BR},x_{CR}),E_C(x_{BL},x_{CL}),E_C(x_{BL},x_{CR}),E_C(x_{BR},x_{CL}),E_C(x_{BR},x_{CR}) ] = $
 = \normalsize
\begin{eqnarray}
 \begin{bmatrix}
         E_{pA,L} & 0          & 0  & 0   & t_{sA}e^{+i\Theta_{P,A}} & 0 & 0 & 0 \\
                0 &   E_{pA,L} & 0        & 0                        & 0                        & t_{sA}e^{+i\Theta_{P,A}} & 0 & 0 \\
                0 & 0          & E_{pA,L} & 0 & 0 & 0 & t_{sA}e^{+i\Theta_{P,A}} & 0 \\
                0 & 0          & 0        &  E_{pA,L} & 0 & 0 & 0 & t_{sA}e^{+i\Theta_{P,A}} \\
                t_{sA}e^{-i\Theta_{P,A}} & 0 & 0 & 0 &  E_{pA,R} & 0 & 0 & 0 \\
                0 & t_{sA}e^{-i\Theta_{P,A}} & 0 & 0 & 0 0 &  E_{pA,R}  & 0 & 0 \\
                0 & 0 & t_{sA}e^{-i\Theta_{P,A}} & 0 & 0 & 0 &  E_{pA,R}  & 0 \\
                0 & 0 & 0 & t_{sA}e^{-i\Theta_{P,A}} & 0 & 0 & 0 &   E_{pA,R} \\
    \end{bmatrix} 
    + \nonumber \\
    +
    \begin{bmatrix}
         E_{pB,L} & 0 & t_{sB}e^{+i\Theta_{P,B}} & 0 & 0 & 0 & 0 & 0 \\
        0 &   E_{pB,L} & 0 & t_{sB}e^{+i\Theta_{P,B}} & 0 & 0 & 0 & 0 \\
        t_{sB}e^{-i\Theta_{P,B}} & 0 & E_{pB,R} & 0 & 0 & 0 & 0 & 0 \\
        0 & t_{sB}e^{-i\Theta_{P,B}} & 0 &  E_{pB,R} & 0 & 0 & 0 & 0 \\
        0 & 0 & 0 & 0 &  E_{pB,L} & 0 & t_{sB}e^{+i\Theta_{P,B}} & 0 \\
        0 & 0 & 0 & 0 & 0  &  E_{pB,L}  & 0 & t_{sB}e^{+i\Theta_{P,B}} \\
        0 & 0 & 0 & 0 & t_{sB}e^{-i\Theta_{P,B}} & 0 &  E_{pB,R}  & 0 \\
        0 & 0 & 0 & 0 & 0 & t_{sB}e^{-i\Theta_{P,B}} & 0 &   E_{pB,R} \\
    \end{bmatrix} 
+ \nonumber \\
+
      \begin{bmatrix}
         E_{pC,L} & t_{sC}e^{+i\Theta_{P,C}} & 0 & 0 & 0 & 0 & 0 & 0 \\
        t_{sC}e^{-i\Theta_{P,C}} &   E_{pC,R} & 0 & 0 & 0 & 0 & 0 & 0 \\
        0 & 0 & E_{pC,L} & t_{sC}e^{+i\Theta_{P,C}} & 0 & 0 & 0 & 0 \\
        0 & 0 & t_{sC}e^{-i\Theta_{P,C}} &  E_{pC,R} & 0 & 0 & 0 & 0 \\
        0 & 0 & 0 & 0 &  E_{pC,L} & t_{sC}e^{+i\Theta_{P,C}} & 0 & 0 \\
        0 & 0 & 0 & 0 & t_{sC}e^{-i\Theta_{P,C}} &  E_{pC,R}  & 0 & 0 \\
        0 & 0 & 0 & 0 & 0 & 0 &  E_{pC,L}  & t_{sC}e^{+i\Theta_{P,C}} \\
        0 & 0 & 0 & 0 & 0 & 0 & t_{sC}e^{-i\Theta_{P,C}} &   E_{pC,R} \\
    \end{bmatrix} 
    +
\end{eqnarray}
\normalsize
\small
$ + diag[ E_C(x_{AL},x_{BL}),E_C(x_{AL},x_{BL}), E_C(x_{AL},x_{BR}),E_C(x_{AL},x_{BR}),E_C(x_{AR},x_{BL}),E_C(x_{AR},x_{BL}),E_C(x_{AR},x_{BL}),E_C(x_{AR},x_{BL}) ] $
$
+diag(E_C(x_{LA},x_{CL}),E_C(x_{LA},x_{CR}),E_C(x_{LA},x_{CL}],E_C(x_{LA},x_{CR}),E_C(x_{RA},x_{CL}),E_C(x_{RA},x_{CR}),E_C(x_{RA},x_{CL}),E_C(x_{RA},x_{CR}) ]$
$ + diag [ E_C(x_{BL},x_{CL}),E_C(x_{BL},x_{CR}), E_C(x_{BR},x_{CL}),E_C(x_{BR},x_{CR}),E_C(x_{BL},x_{CL}),E_C(x_{BL},x_{CR}),E_C(x_{BR},x_{CL}),E_C(x_{BR},x_{CR}) ] . $
  \normalsize
\subsection{Electrostatic interaction between four position-based qubits }

\begin{figure}
    \centering
    \begin{subfigure}[b]{0.2\textwidth}
         \centering
         \includegraphics[width=\textwidth]{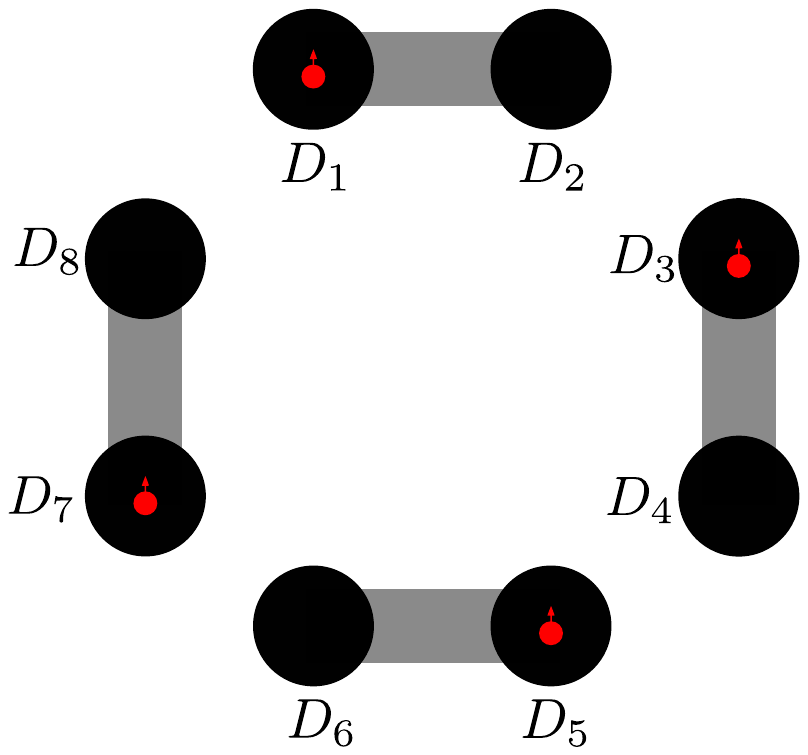}
         \caption{$t_1$}
         \label{fig:QDt1}
     \end{subfigure}
     \hfill
     \begin{subfigure}[b]{0.2\textwidth}
         \centering
         \includegraphics[width=\textwidth]{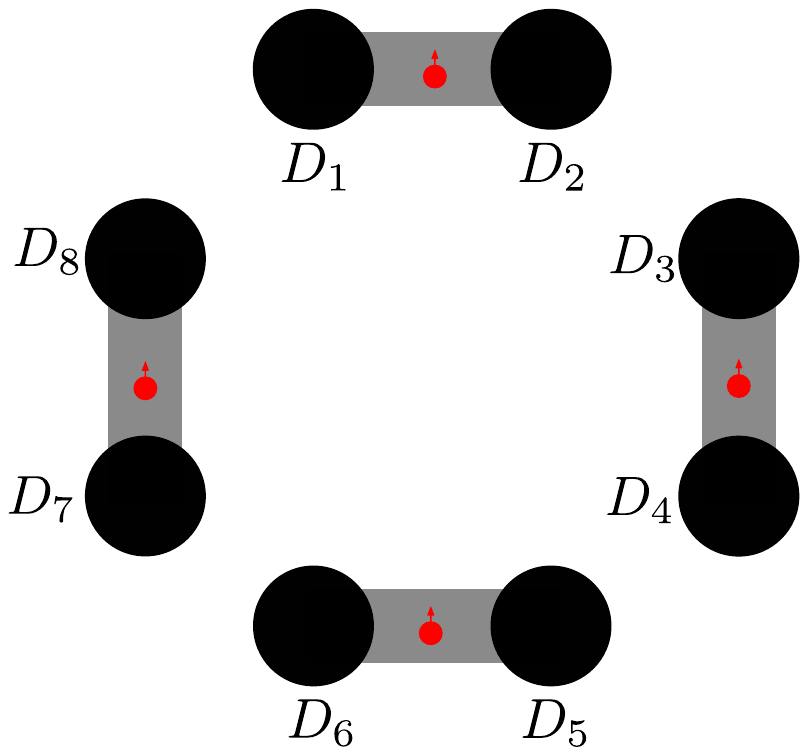}
         \caption{$t_2$}
         \label{fig:QDt2}
     \end{subfigure}
     \hfill
     \begin{subfigure}[b]{0.2\textwidth}
         \centering
         \includegraphics[width=\textwidth]{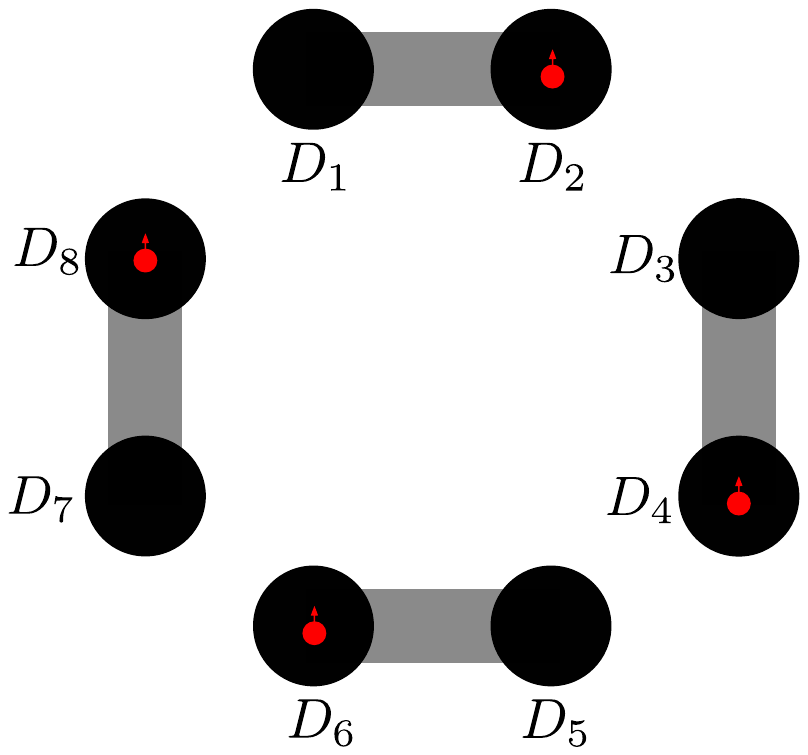}
         \caption{$t_3$}
         \label{fig:QDt3}
     \end{subfigure}
     \hfill
     \begin{subfigure}[b]{0.2\textwidth}
         \centering
         \includegraphics[width=\textwidth]{images/QD3.pdf}
         \caption{$t_4$}
         \label{fig:QDt4}
     \end{subfigure}
        \caption{Scheme of four electrostatically interacting single-electron position based qubits at different stages of interaction, when $t_1<t_2<t_3<t_4$}
        \label{fig:Qdots}
\end{figure}

\begin{figure}
    \centering
    \includegraphics[width=0.5\linewidth]{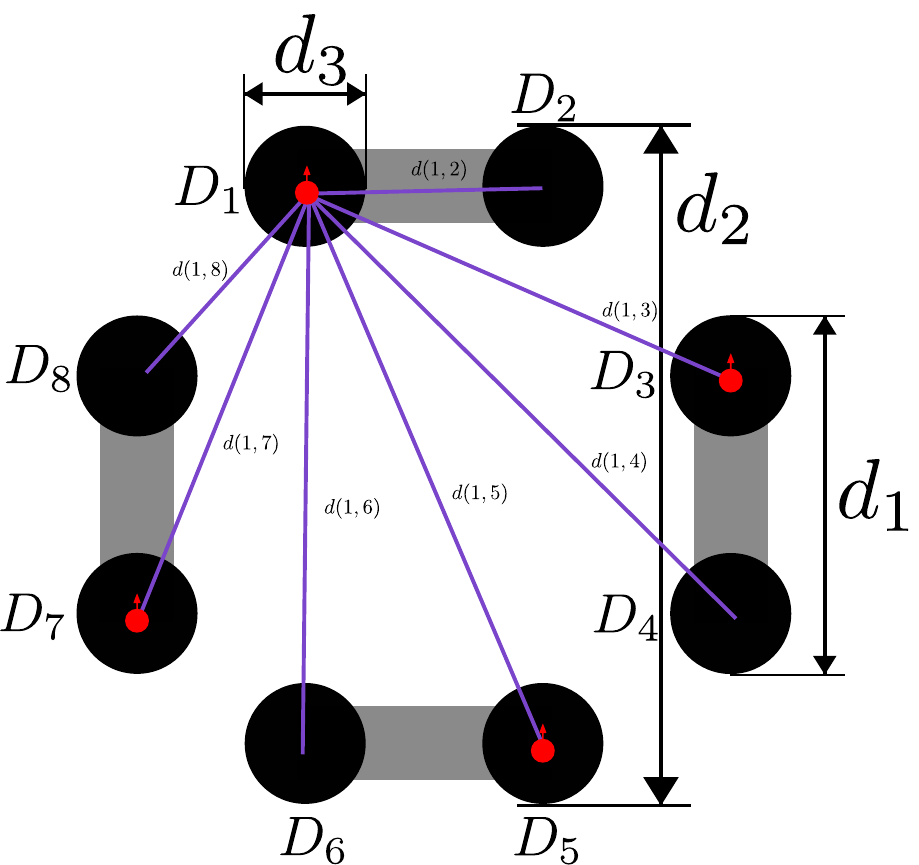}
    \caption{Scheme of 4 electrostatically interacting position-based qubits with rotation symmetry $\frac{\pi}{2}$.}
    \label{fig:DistQD}
\end{figure}

For a system of four quantum dots such as that presented in Figure \ref{fig:Qdots}, the interaction can be written as:
\begin{multline}
\label{eq:HInt4x4}
    \hat{H}_{Int} = E_C(D_1,D_3) \ket{x_{D1}}\bra{x_{D1}} \otimes_K \ket{x_{D3}}\bra{x_{D3}} \otimes_K \hat{I} \otimes_K \hat{I} +
                    E_C(D_1,D_4) \ket{x_{D1}}\bra{x_{D1}} \otimes_K \ket{x_{D4}}\bra{x_{D4}} \otimes_K \hat{I} \otimes_K \hat{I} +\\
                    E_C(D_2,D_3) \ket{x_{D2}}\bra{x_{D2}} \otimes_K \ket{x_{D3}}\bra{x_{D3}} \otimes_K \hat{I} \otimes_K \hat{I} +
                    E_C(D_2,D_4) \ket{x_{D2}}\bra{x_{D2}} \otimes_K \ket{x_{D4}}\bra{x_{D4}} \otimes_K \hat{I} \otimes_K \hat{I} +\\
                    E_C(D_1,D_5) \ket{x_{D1}}\bra{x_{D1}} \otimes_K \hat{I} \otimes_K \ket{x_{D5}}\bra{x_{D5}} \otimes_K \hat{I} +
                    E_C(D_1,D_6) \ket{x_{D1}}\bra{x_{D1}} \otimes_K \hat{I} \otimes_K \ket{x_{D6}}\bra{x_{D6}} \otimes_K \hat{I} +\\
                    E_C(D_2,D_5) \ket{x_{D2}}\bra{x_{D2}} \otimes_K \hat{I} \otimes_K \ket{x_{D5}}\bra{x_{D5}} \otimes_K \hat{I} +
                    E_C(D_2,D_6) \ket{x_{D2}}\bra{x_{D2}} \otimes_K \hat{I} \otimes_K \ket{x_{D6}}\bra{x_{D6}} \otimes_K \hat{I} +\\
                    E_C(D_1,D_7) \ket{x_{D1}}\bra{x_{D1}} \otimes_K \hat{I} \otimes_K \hat{I} \otimes_K \ket{x_{D7}}\bra{x_{D7}} +
                    E_C(D_1,D_8) \ket{x_{D1}}\bra{x_{D1}} \otimes_K \hat{I} \otimes_K \hat{I} \otimes_K \ket{x_{D8}}\bra{x_{D8}} +\\
                    E_C(D_2,D_7) \ket{x_{D2}}\bra{x_{D2}} \otimes_K \hat{I} \otimes_K \hat{I} \otimes_K \ket{x_{D7}}\bra{x_{D7}} +
                    E_C(D_2,D_8) \ket{x_{D2}}\bra{x_{D2}} \otimes_K \hat{I} \otimes_K \hat{I} \otimes_K \ket{x_{D8}}\bra{x_{D8}} +\\
                    \hat{I} \otimes_K E_C(D_3,D_5) \ket{x_{D3}}\bra{x_{D3}} \otimes_K \ket{x_{D5}}\bra{x_{D5}} \otimes_K \hat{I} +
                    \hat{I} \otimes_K E_C(D_3,D_6) \ket{x_{D3}}\bra{x_{D3}} \otimes_K \ket{x_{D6}}\bra{x_{D6}} \otimes_K \hat{I} +\\
                    \hat{I} \otimes_K E_C(D_4,D_5) \ket{x_{D4}}\bra{x_{D4}} \otimes_K \ket{x_{D5}}\bra{x_{D5}} \otimes_K \hat{I} +
                    \hat{I} \otimes_K E_C(D_4,D_6) \ket{x_{D4}}\bra{x_{D4}} \otimes_K \ket{x_{D6}}\bra{x_{D6}} \otimes_K \hat{I} +\\
                    \hat{I} \otimes_K E_C(D_3,D_7) \ket{x_{D3}}\bra{x_{D3}} \otimes_K \hat{I} \otimes_K \ket{x_{D7}}\bra{x_{D7}} +
                    \hat{I} \otimes_K E_C(D_3,D_8) \ket{x_{D3}}\bra{x_{D3}} \otimes_K \hat{I} \otimes_K \ket{x_{D8}}\bra{x_{D8}} +\\
                    \hat{I} \otimes_K E_C(D_4,D_7) \ket{x_{D4}}\bra{x_{D4}} \otimes_K \hat{I} \otimes_K \ket{x_{D7}}\bra{x_{D7}} +
                    \hat{I} \otimes_K E_C(D_4,D_8) \ket{x_{D4}}\bra{x_{D4}} \otimes_K \hat{I} \otimes_K \ket{x_{D8}}\bra{x_{D8}} +\\
                    \hat{I} \otimes_K \hat{I} \otimes_K E_C(D_5,D_7) \ket{x_{D5}}\bra{x_{D5}} \otimes_K \ket{x_{D7}}\bra{x_{D7}} +
                    \hat{I} \otimes_K \hat{I} \otimes_K E_C(D_5,D_8) \ket{x_{D5}}\bra{x_{D5}} \otimes_K \ket{x_{D8}}\bra{x_{D8}} +\\
                    \hat{I} \otimes_K \hat{I} \otimes_K E_C(D_6,D_7) \ket{x_{D6}}\bra{x_{D6}} \otimes_K \ket{x_{D7}}\bra{x_{D7}} +
                    \hat{I} \otimes_K \hat{I} \otimes_K E_C(D_6,D_8) \ket{x_{D6}}\bra{x_{D6}} \otimes_K \ket{x_{D8}}\bra{x_{D8}}
\end{multline}
where $E_C(a,b)$ is the Coulomb interaction between nodes $a$ and $b$, $\ket{x_{Da}}\bra{x_{Da}}$ denotes:
\begin{gather}
    \begin{cases}
        \ket{x_{Da}}\bra{x_{Da}} = \begin{bmatrix}1&0\\0&0\end{bmatrix} = \left( \frac{1}{2}\hat{\sigma_0}+\frac{1}{2}\hat{\sigma_1} \right), & \forall_{a\in\mathbb{N}\land a\in[1,8]} a\bmod2 \equiv 1  \\
        \ket{x_{Da}}\bra{x_{Da}} = \begin{bmatrix}0&0\\0&1\end{bmatrix} = \left( \frac{1}{2}\hat{\sigma_0}-\frac{1}{2}\hat{\sigma_1} \right), & \forall_{a\in\mathbb{N}\land a\in[1,8]} a\bmod2 \equiv 0 \\
    \end{cases}
\end{gather}
The Coulomb therm $E_C(a,b)$ in explicit form looks as follows:
\begin{equation}
    E_C(a,b) = \frac{e^2}{d(a,b)}
\end{equation}
where $d(a,b)$ is distance between points $a$ and $b$. For structure similar to this on figure \ref{fig:DistQD} we can directly denote $d(a,b)$ to:
\begin{equation}
    \begin{cases}
        d(1,2) = d_1 - d_3\\
        d(1,3) = \sqrt{\left( \frac{d_2 - d_1}{2} \right)^2 + \left( d_1 - \frac{d_3}{2} +\frac{d_2 - d_1}{2} \right)^2}\\
        d(1,4) = \sqrt{\left( d_1 - d_3 +\frac{d_2 - d_1}{2}\right)^2 + \left( d_1 - \frac{d_3}{2} +\frac{d_2 - d_1}{2}\right)^2}\\
        d(1,5) = \sqrt{\left(d_1 - d_3 \right)^2 + d_2^2}\\
        d(1,6) = d_2 \\
        d(1,7) = \sqrt{\left(\frac{d_2 -d_1}{2}\right)^2 + \left(\frac{d_2 -d_1}{2}\right)^2}\\
        d(1,8) = \sqrt{\left(\frac{d_2 -d_1}{2}\right)^2 + \left(\frac{d_2 -d_1}{2} + d_1 - d_3\right)^2}
    \end{cases}
\end{equation}
Using the generalized Kronecker double sum, we can reduce the equation \eqref{eq:HInt4x4} to:
\begin{multline}
    \hat{H}_{Int} = \chi_{1,n+1}^{3,4}\left( \frac{1}{2}(\hat{\sigma}_0+\hat{\sigma}_1),\frac{1}{2}(\hat{\sigma}_0+\hat{\sigma}_1), E_C(2n-1,2m-1) \right) + \\ + \chi_{1,n+1}^{3,4}\left(\frac{1}{2}(\hat{\sigma}_0+\hat{\sigma}_1),\frac{1}{2}(\hat{\sigma}_0-\hat{\sigma}_1), E_C(2n-1,2m) \right) + \\ + \chi_{1,n+1}^{3,4}\left(\frac{1}{2}(\hat{\sigma}_0-\hat{\sigma}_1),\frac{1}{2}(\hat{\sigma}_0+\hat{\sigma}_1), E_C(2n,2m-1) \right) +\\+ \chi_{1,n+1}^{3,4}\left(\frac{1}{2}(\hat{\sigma}_0-\hat{\sigma}_1),\frac{1}{2}(\hat{\sigma}_0-\hat{\sigma}_1), E_C(2n,2m) \right)
\end{multline}
Finally the Hamiltonian for system of four interacting position based qubits can be described with the following sum of generalized Kronecker double sums:
\begin{multline}
    \hat{H} = \chi_{1,4}^{4,4}\left( \hat{\sigma}_0, \hat{I}, \langle E_n \rangle \right) + \chi_{1,4}^{4,4}\left( \vb*{\hat{q}_n}, \hat{I}, \text{i} \right) \\ + \chi_{1,n+1}^{3,4}\left( \frac{1}{2}(\hat{\sigma}_0+\hat{\sigma}_1),\frac{1}{2}(\hat{\sigma}_0+\hat{\sigma}_1), E_C(2n-1,2m-1) \right) + \\ + \chi_{1,n+1}^{3,4}\left(\frac{1}{2}(\hat{\sigma}_0+\hat{\sigma}_1),\frac{1}{2}(\hat{\sigma}_0-\hat{\sigma}_1), E_C(2n-1,2m) \right) + \\ + \chi_{1,n+1}^{3,4}\left(\frac{1}{2}(\hat{\sigma}_0-\hat{\sigma}_1),\frac{1}{2}(\hat{\sigma}_0+\hat{\sigma}_1), E_C(2n,2m-1) \right) +\\+ \chi_{1,n+1}^{3,4}\left(\frac{1}{2}(\hat{\sigma}_0-\hat{\sigma}_1),\frac{1}{2}(\hat{\sigma}_0-\hat{\sigma}_1), E_C(2n,2m) \right)
\end{multline}
\subsection{Conclusions and further perspectives}
Explicit schemes of quaternionic description for two-energy-level Hamiltonian for position-based qubits are given in the quaternionic representation.
In addition, the quantum evolution operator in quaternionic form in the case of a single quantum body (position-based qubit) and with the situation of multiple interacting quantum bodies (case of many interacting qubits) is presented. Quantum state expressed by density matrix multiplied by imaginary unit turns out to be quaternionic and is given by equation of motion in real-value time as pointed by formulas \ref{eq:0}, \ref{eq:1}, \ref{eq:2}, \ref{eq:3}, \ref{eq:4}. 
It is quite straightforward to extend the presented scheme to quantum N-body weakly electrostatically interacting qubits. 
Cases of 2 , 3 and 4 interacting bodies were described explicitly. 
Both authors contributed equally in 50 percent to this work. The first author contribution is due to technical part of the work, while the second author contribution is mostly conceptual. 

%We have \cite{IBMQExp}, \cite{10.1063/1.1703794}
%%%\printbibliography

\end{document}